\documentstyle[aps,epsfig]{revtex}

\begin{document}
\draft 
\preprint{} 
\title{Local and global properties of
conformally flat initial data for black hole collisions} \makeatletter
\author{Nina Jansen} 
\address{Theoretical Astrophysics Center,Denmark, jansen@tac.dk} 
\author{Peter Diener}
\address{Max-Planck-Institut f$\ddot{u}$r Gravitationsphysik,
Albert-Einstein-Institut, Germany, diener@aei-potsdam.mpg.de}
\author{Alexei Khokhlov} 
\address{Code 6404, Naval Research Laboratory, Washington DC, 
ajk@lcp.nrl.navy.mil} 
\author{Igor Novikov} 
\address{Theoretical Astrophysics Center\cite{byline}, 
Denmark, novikov@tac.dk}
\date{\today} 
\maketitle 
\makeatother
\begin{abstract}
We study physical
properties of conformal initial value data for single and binary black
hole configurations obtained using conformal-imaging and
conformal-puncture methods. We investigate how the total mass
$M_{tot}$ of a dataset with two black holes depends on the
configuration of linear or angular momentum and separation of the
holes. The asymptotic behavior of $M_{tot}$ with increasing separation
allows us to make conclusions about an unphysical ``junk'' gravitation
field introduced in the solutions by the conformal approaches. We also
calculate the spatial distribution of scalar invariants of the Riemann
tensor which determine the gravitational tidal forces. For single
black hole configurations, these are compared to known analytical
solutions. Spatial distribution of the invariants allows us to make
certain conclusions about the local distribution of the additional
field in the numerical datasets.
\end{abstract}
\widetext

\section{Introduction}
\label{sec:1}
\setlength{\parindent}{0em} \setlength{\parskip}{1.5ex plus 0.5ex}

The problem of initial conditions (initial Cauchy data) for the
integration of the evolution of colliding black holes is an important
problem of the numerical general relativity and has attracted
attention of many researchers.  For a recent review see
\cite{Cook2000}. One of the approaches to the problem is a
conformal-imaging method proposed in \cite{York3}, \cite{York1},
\cite{York2} and developed in \cite{Cook1}, \cite{Cook2}. Another is
the conformal-puncture method \cite{BB}. It is known that initial
black hole data constructed using both these methods contains some
non-vanishing dynamical component \cite{Cook2000}, an unphysical, junk
gravitational field.  Without the integration of the initial data in
time it is impossible to make exact conclusions about the character
and the amount of this unphysical field.  Still it is possible to make
some conclusions about the junk field using only the initial data. For
example, one can numerically calculate global characteristics of the
initial data such as total energy, and then compare it with known
analytical or approximate solutions
\cite{Cook1},\cite{Cook2},\cite{BB}.

In our previous paper we developed adaptive mesh refinement approach
to the construction of initial data for black hole collisions on high
resolution Cartesian meshes \cite{DJKN1}. The method allows us to
compute initial data with high accuracy both near and very far away
the from black holes. The goal of this paper is to use this method to 
systematically analyze the physical properties of black hole initial data
for a wide range of colliding black hole configurations.  To analyze
the junk fields present in the data, we will use two different
approaches. First, we calculate and compare global characteristics of
the configurations. In addition, we use a new approach which consists
of calculating and comparing four local scalar invariants of the
gravitational field (\cite{LL}, section 92) . These scalar invariants
completely determine local properties (tidal forces) of a
gravitational field. 

In the next section \ref{sec:2} we briefly describe our adaptive mesh
refinement method of calculating initial black hole data.  We start
our discussion of local and global characterization of the
solutions in section \ref{sec:3}.  Results for single black hole
configurations with angular or linear momenta are presented in section
\ref{sec:4}. For these black holes we can compare numerical initial
date with exact analytical solutions. In section \ref{sec:5} we
present results for two-black hole configurations with different
orientations of linear and angular momenta and with different
separations between the black holes. All calculations are performed
using both conformal-imaging and conformal-puncture methods, and the
results are compared. Conclusions are presented in section
\ref{sec:6}.

\section{Constraint equations and the method of solution}
\label{sec:2}
\setlength{\parindent}{0em} \setlength{\parskip}{1.5ex plus 0.5ex}

The ADM or 3+1 formulation of the equations of general relativity
works with the metric $g^{ph}_{ij}$ and extrinsic curvature
$K^{ph}_{ij}$ of three-dimensional spacelike hypersurfaces embedded in
the four-dimensional space-time, where $i,j = 1,2,3$, and the
superscript $ph$ denotes the physical space. On the initial
hypersurface, $g^{ph}_{ij}$ and $K^{ph}_{ij}$ must satisfy the
constraint equations \cite{York3}. The conformal approach assumes that
the metric is conformally flat,
\begin{equation}
\label{eq:1}
g^{ph}_{ij} = \phi^4 g_{ij} ~,
\end{equation}
where $g_{ij}$ is the metric of a background flat space.  This
conformal transformation induces the corresponding transformation of
the extrinsic curvature
\begin{equation}
\label{eq:2}
K^{ph}_{ij} = \phi^{-2} K_{ij}~.
\end{equation}
With the additional assumption of
\begin{equation}
\label{eq:3}
tr K = 0~,
\end{equation}
the energy and momentum constraints are
\begin{equation}
\label{eq:4}
\nabla^2 \phi + \frac{1}{8} \phi^{-7} K_{ij} K^{ij} = 0,
\end{equation}
and
\begin{equation}
\label{eq:5}
D_j K^{ij} = 0,
\end{equation}
respectively, where $\nabla^2$ and $D_j$ are the Laplacian and
covariant derivative in flat space.

A solution to (\ref{eq:4}) and (\ref{eq:5}) for two black holes can be
specified by six parameters (three parameters per black hole). These
are the mass parameter $M_\delta$, linear momentum parameter
$\vec{P}_\delta$ and angular momentum parameters $\vec{S}_\delta$, where
$\delta=1,2$ is the index of a black hole. In terms of these
parameters, the solution to the momentum constraint for two black
holes is
\begin{equation}
\label{eq:6}
K_{ij}({\bf r}) = K^{lin}_{ij}({\bf r}) + K^{ang}_{ij}({\bf r}) ~,
\end{equation}
where
\begin{equation}
\label{eq:7}
    K^{lin}_{ij}({\bf r}) =
   \frac{3}{2} \sum_{\delta=1}^2
  \frac{1}{R_\delta^2} \left( P_{\delta,i} n_{\delta,j} +
   P_{\delta,j} n_{\delta,i} -
   \left( g_{ij} - n_{\delta,i} n_{\delta,j} \right)
   P_{\delta,k} n_\delta^k
   \right) 
\end{equation}
and
\begin{equation}
\label{eq:8}
K^{ang}_{ij}({\bf r}) =
3 \sum_{\delta=1}^2 \left(
\frac{1}{R_\delta^3}
\left( \epsilon_{kil} S_\delta^l n_\delta^k n_{\delta,j} +
\epsilon_{kjl} S_\delta^l n_\delta^k n_{\delta,i} \right)
\right) \quad.
\end{equation}
In (\ref{eq:6}) and (\ref{eq:7}) the comma separates the index of
black hole from the coordinate component indices,
$R_\delta=M_\delta/2$ is the black hole throat radius, and ${\bf
n}_\delta = ( {\bf r} - {\bf r}_\delta ) / \vert {\bf r} - {\bf
r}_\delta \vert$ is the unit vector directed from the center of the
$\delta$-th black hole ${\bf r}_\delta$ to the point $\bf r$. We work
in units where $G=1,~c=1$ which means that $\vec{P}_\delta$ is in mass
units and $\vec{S}_\delta$ is in mass$^2$ units. 

In the conformal-imaging method, we obtain an inversion-symmetric
solution to (\ref{eq:5}) by applying an infinite series of mirror
operators to (\ref{eq:6}), as described in \cite{Cook1}.  Note, that
before applying the mirror operators to $K^{ang}_{ij}$, this term must
be divided by 2 since the image operators will double its value.  The
series converges rapidly, and in practice only a few terms are
taken. After the isometric solution for $K_{ij}$ is found, (\ref{eq:4})
must be solved subject to the isometry boundary condition at the black
hole throats
\begin{equation}
\label{eq:9}
n_\delta^i D_i \phi = -\frac{\phi}{2 R_\delta}~.
\end{equation}

In the conformal-puncture method, expression (\ref{eq:6}) is used
without mirror imaging, and the energy constraint is solved on $R^3$
with a puncture at the center of each hole. Brandt and
Br$\ddot{u}$gmann \cite{BB} proved that it suffices to solve the
energy constraint everywhere on $R^3$ without any points cut
out. Thus, this method avoids inner boundaries.

The outer boundary condition in both cases is $\phi \rightarrow 1$ at
infinity. This boundary condition is represented by \cite{Cook1}
\begin{equation}
\label{eq:10}
\frac{\partial \phi}{\partial r} =\frac{1-\phi}{r} ~,
\end{equation}
where $r$ is the distance from the center of the computational domain to the
boundary.

As described in \cite{DJKN1}, (\ref{eq:4}) and (\ref{eq:5}) are solved
on an adaptive Cartesian mesh.  The mesh is refined on the level of
individual cells using a fully threaded tree structure
\cite{FTT}. Computational cells can be characterized by the level of
cells $l$ in the tree.  A cell size $\Delta_l$ is related to the cell
level as $\Delta_l = L / 2^l$, where $L$ is the size of the
computational domain. In this paper, $L=1024$ and the minimum and
maximum levels of refinement is $l_{min}=5$ and $l_{max}=16$,
respectively. That is, the coarsest cells had a size $\Delta_{min}
=32$ and the finest cell size was $\Delta_{max}= 1/64$. The unit of
length is the throat radius $R_\delta=M_\delta/2$ of the holes (in
this paper, all holes have the same mass $M_\delta = 2$ and thus the
same throat radius).  The equivalent uniform grid resolution thus
corresponded to that of a $65536^3$ uniform grid.  Mesh was refined at
$l_{min} < l < l_{max}$ according to the following refinement
criterium (see \cite{DJKN1} for details)
\begin{equation}
\label{eq:11}
\eta = \max \left(
      \frac{\Delta}{\phi^4} \left(
      \left(\frac{\partial\phi^4}{\partial x}\right)^2 +
      \left(\frac{\partial\phi^4}{\partial y}\right)^2 +
      \left(\frac{\partial\phi^4}{\partial z}\right)^2
                              \right)^{1/2}, \vert K_{ij}
      \right)~.
\end{equation}
Mesh was refined at levels $l=5,6$ if $\eta < \epsilon=0.03$ and at
$l> 6$ if $\eta < \epsilon=0.07$. In our previous paper it was
demonstrated that our mesh refinement approach gives a quadratic
convergence of solutions with increasing numerical resolution. In this
paper we additionally checked the accuracy of numerical solutions by
varying $\epsilon$, $l_{max}$ and $l_{min}$.  The accuracy of the
results is discussed further in sections \ref{sec:4} and
\ref{sec:5}.

\section{Global and local characteristics of initial data}
\label{sec:3}
\setlength{\parindent}{0em} \setlength{\parskip}{1.5ex plus 0.5ex}
Global physical properties of a system are its total mass $M_{tot}$,
linear, ${\bf P}_{tot}$, and angular, ${\bf S}_{tot}$, momentum
\cite{York4}. For conformally flat solutions
\begin{equation}
\label{eq:12}
M_{tot} =
-\frac{1}{2 \pi} \oint_{r\rightarrow\infty} (\frac{\partial \phi}{\partial x^j})
d\sigma^j~,
\end{equation}
\begin{equation}
\label{eq:13}
P_{i,tot} =\frac{1}{8 \pi}\oint_{r \rightarrow \infty}
K_{ij} d\sigma^j ~,
\end{equation}
\begin{equation}
\label{eq:14}
S_{i,tot} = \frac{1}{8 \pi}\varepsilon_{kij}
\oint_{r \rightarrow \infty}  (x^i K^{jk} - x^j K^{ik}) d\sigma_k~,
\end{equation}
where integration is performed over the surface at infinity. For our
case of two black holes, the total linear and angular momenta are equal to
\begin{equation}
\label{eq:15}
{\bf P}_{tot} = {\bf P}_1 + {\bf P}_2 ~,
\end{equation}
and
\begin{equation}
\label{eq:16}
{\bf S}_{tot} = {\bf S}_1 + {\bf S}_2 +
({\bf C}_1 - {\bf O}) \times {\bf P}_1 + ({\bf C}_2 -{\bf O}) \times 
{\bf P}_2 ~,
\end{equation}
where ${\bf C}_1,{\bf C}_2$ are centers of black hole throats or
punctures and ${\bf O}$ is the center of mass of the system (see
\cite{Cook4}).  Note, that ${\bf P}_1, {\bf P}_2, {\bf S}_1, {\bf S}_2$
do not have direct physical meaning but ${\bf P}_{tot}$ and ${\bf
S}_{tot}$ do.

In addition to global characteristics of a system, we can also
consider local invariants.  In general case, there are four such
invariants (\cite{LL}, section 92)
\begin{eqnarray}
\label{eq:17}
I_1 &=& R^{\alpha\beta\gamma\delta}R_{\alpha\beta\gamma\delta}\\
I_2 &=& R^{\alpha\beta\gamma\delta}\hat R_{\alpha\beta\gamma\delta}\\
I_3 &=& R_{\alpha\beta\gamma\delta}R^{\gamma\delta\mu\nu}
R_{\mu\nu}^{\alpha\beta}\\
I_4 &=& R_{\alpha\beta\gamma\delta}R^{\gamma\delta\mu\nu}
\hat R_{\mu\nu}^{\alpha\beta}
\end{eqnarray}
where $ \hat R_{\alpha\beta\gamma\delta} = \frac{1}{2}
\varepsilon_{\alpha\beta\mu\nu} R^{\mu\nu}_{\gamma\delta}$ and
$\varepsilon_{\alpha\beta\mu\nu}$ is a fully antisymmetric unit tensor
in curved coordinates. At every point, in a specially selected
tetrad, the Riemann tensor can be expressed in terms of these
invariants (except of special degenerate cases \cite{LL}). Since
Riemann tensor uniquely determines tidal forces, these invariants
fully characterize physical properties of the field at every
point. The tidal force is $\propto I_1^{1/2}$ by the order of
magnitude. For a Schwarzschild black hole, for example, the tidal force
is exactly $M/c^2 r^3 = \sqrt{I_1}$ \cite{LL}.

In order to calculate the Riemann tensor, we need to know the space
and time derivatives of the four-dimensional metric at each point on
the initial slice.  This data can be obtained from the time-evolution
part of the Einstein equations. Since the scalars do not depend on the
choice of a coordinate system, we can select zero shift and unit lapse
to simplify calculations, and to write the derivatives as
\begin{eqnarray}
\label{eq:18}
\frac{\partial g_{ij}}{\partial t} &=& -2 K_{ij} ~, \\
\frac{\partial^2 g_{ij}}{\partial t \partial x^k} &=& -2
\frac{\partial K_{ij}}{\partial x^k}~,\\
\frac{\partial^2 g_{ij}}{\partial t^2} &=& -2(\,^{(3)}R_{ij} + trK 
K_{ij} -2 K_{ik} K^k_{\;j})~.
\end{eqnarray}
Here both $g_{ij},K_{ij}$ are the physical metric and extrinsic
curvature, and not their counterparts in flat space, and
$^{(3)}R_{ij}$ is the Ricci tensor of 3-dimensional space which
depends only on the spatial metric and it's spatial derivatives.

\section{Single black holes}
\label{sec:4}
\setlength{\parindent}{0em} \setlength{\parskip}{1.5ex plus 0.5ex}

\subsection{\bf \normalfont \normalsize \it Rotating single black hole} 

We consider a single black hole with a non-zero angular momentum ${\bf
S} \neq 0$ and zero linear momentum ${\bf P}=0$, and compare numerical
conformal solutions with the known analytical solution for a Kerr
black hole.  The Kerr metric is
\begin{equation}
\label{eq:19}
ds^2 = -\left(1 - \frac{2 M_{tot} r}{\Sigma}\right) dt^2
- \frac{4 M_{tot} r a \sin^2\theta}{\Sigma} dt d\phi
+ \frac{\Sigma}{\Delta} dr^2 + \Sigma d\theta^2 +
\frac{A \sin^2\theta}{\Sigma}d\phi^2 ~, 
\end{equation}
where
\begin{eqnarray}
\label{eq:20}
\Sigma &=& r^2 a^2 \cos^2 \theta ~,\\
\Delta &=& r^2 - 2 M_{tot} r a^2 ~, \\
A &=& (r^2 + a^2)^2 - a^2 \Delta \sin^2\theta ~.
\end{eqnarray}
The analytical solution (19),(20) is described by two parameters, the
total mass of the hole $M_{tot}$ measured at infinity, and the
specific angular momentum $a=S/M_{tot}$ such that $a \leq M_{tot}$.
The irreducible mass of a Kerr black hole, i.e., the mass of the event
horizon of the hole is given by
\begin{equation}
\label{eq:21}
M_{irr} = \sqrt{\frac{M_{tot}}{2} \left(M_{tot} +
\sqrt{-\left(\frac{S}{M_{tot}}\right)^2 + M_{tot}^2} \right)} ~. 
\end{equation}

We have calculated datasets with single rotating black holes using
both the imaging and puncture method. In each case, we vary the $S$
parameter, but hold the mass parameter of the black hole, $M=2$,
constant. We have checked the accuracy of the solutions by varying
$l_{max}$, $l_{min}$ and $\epsilon$ in (\ref{eq:11}). For each dataset
we computed $M_{tot}$ and $S_{tot}$ by evaluating (\ref{eq:12}) and
(\ref{eq:14}) over at sphere with radius $r \leq L/2 \simeq 500$. We
found that the values of $M_{tot}$, $S^{tot}_i$ calculated at this
radius were third digit accurate. The integration in (\ref{eq:12}) and
(\ref{eq:14}) is supposed to be over at surface at infinity, and we
determined $M_{tot}(\infty)$ and $S_{tot}(\infty)$ from the
corresponding values at finite radii by extrapolation in $1/r$ to $1/r
= 0$.

\begin{figure}
\begin{center}
\epsfig{file=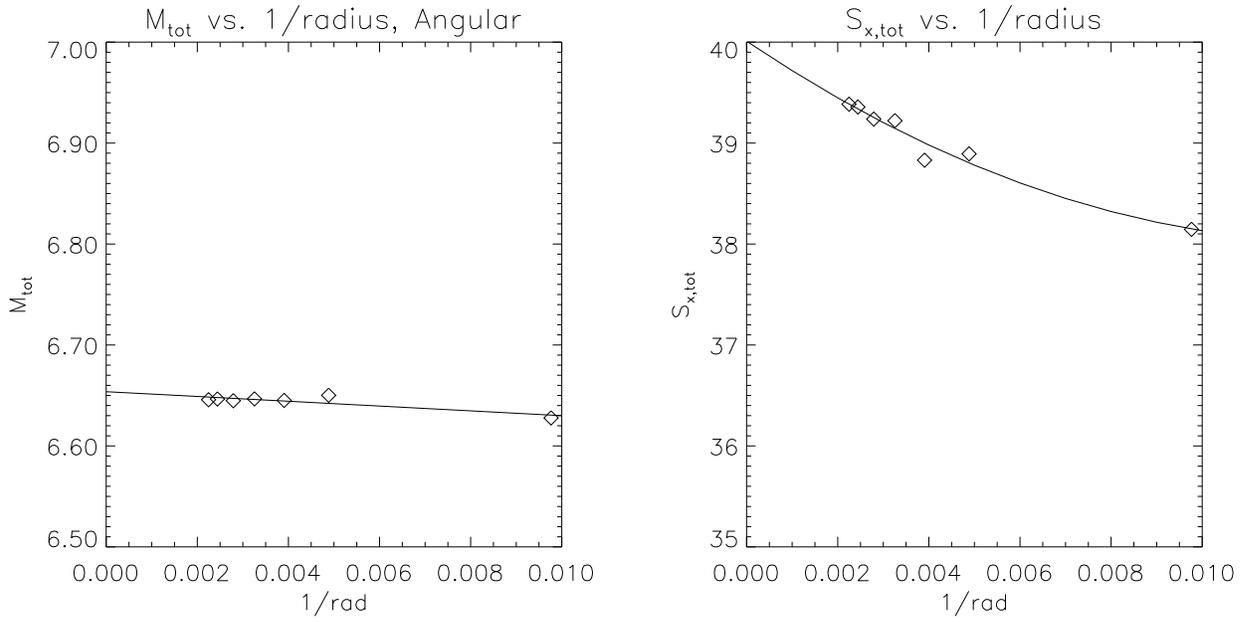,width=17cm}
\caption[]{Asymptotic total mass, $M_{tot}$, and total angular
momentum, $S_{tot}$, for case M=2 S=40.
The diamonds are the numerical values of $M_{tot}$,$S_{tot}$ found
from integrating over a sphere at radius $r$. The line is a 
least square fit of the data to a second order polynomial.
\label{fig:1}}
\end{center}
\end{figure}

\begin{figure}
\begin{center}
\epsfig{file=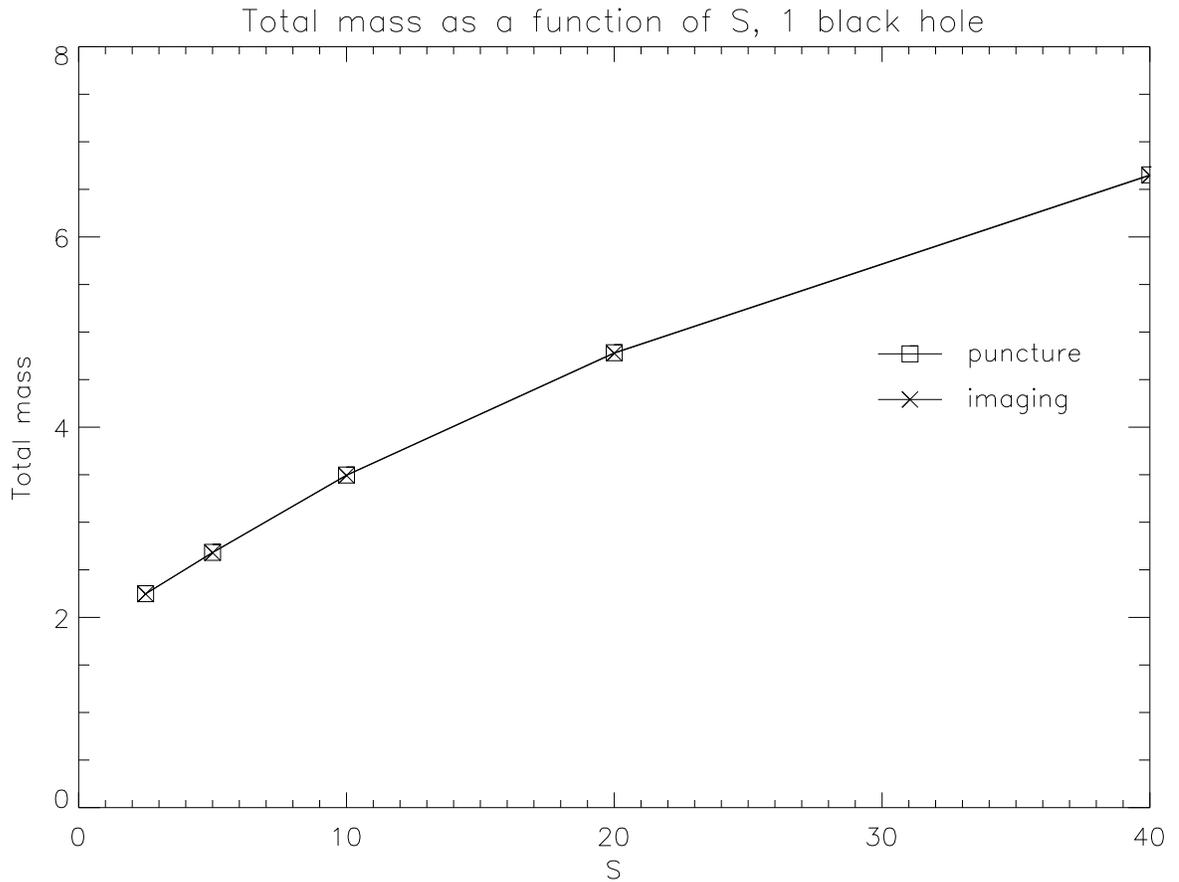,width=17cm}
\caption[]{Total mass, $M_tot$, as a function of S for a single black hole, for
both imaging and puncture method.
\label{fig:2}}
\end{center}
\end{figure}

Figure \ref{fig:1} shows the extrapolation procedure for a rotating
black hole with $S=40$ calculated with the imaging method. $M_{tot}$
and $S_{tot}$ have been calculated by integrating over spheres with
increasing radii, $r$. We find a least square fit of the data to a
second order polynomial and evaluate that polynomial at $r=\infty$ to
obtain $S_{tot}(\infty)$ and $M_{tot}(\infty)$.  For rotating black
hole, the values of $M_{tot}$ are practically not changing with $r$,
whereas the values of $S_{tot}(\infty)$ are slightly larger than the
values at finite $r$. For a case $S=40$, the extrapolated value
$S_{tot}(\infty)\simeq 40 \simeq S$, as it should be. The asymptotic
value of $M_{tot}$ is in a good agreement with that obtained in
\cite{Cook3}. Figure \ref{fig:2} shows $M_{tot}(\infty)$ as a function
of $S$.

\begin{figure}
\begin{center}
\epsfig{file=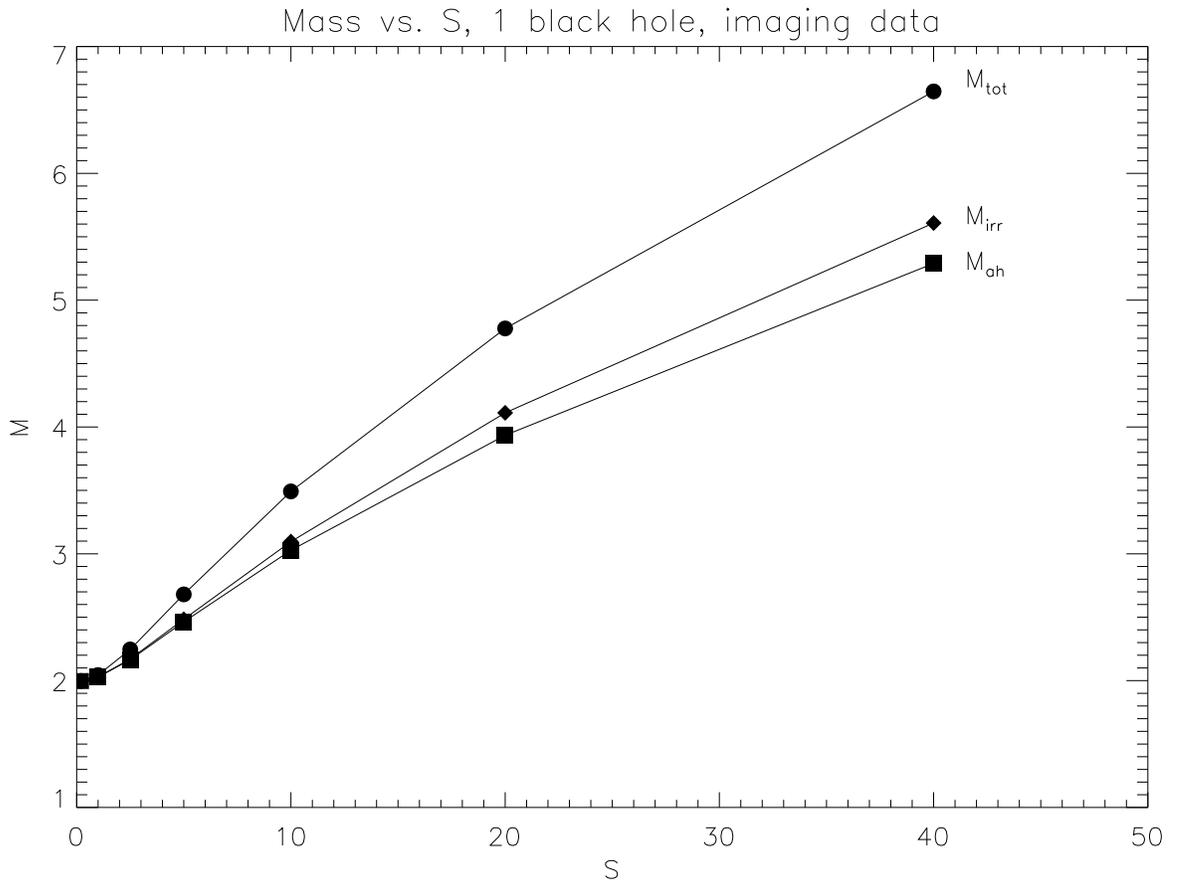,width=17cm}
\caption[]{Total mass, $M_{tot}$, apparent horizon mass, $M_{ah}$, and 
irreducible mass, $M_{irr}$, as a function of S for a 
rotating black hole. $M_{tot}$ and $M_{ah}$ are calculated
numerically, and $M_{irr}$ is the irreducible mass of a Kerr black
hole with the same overall mass and angular momentum as the numerical
black hole.
\label{fig:3}}
\end{center}
\end{figure}

\begin{figure}
\begin{center}
\epsfig{file=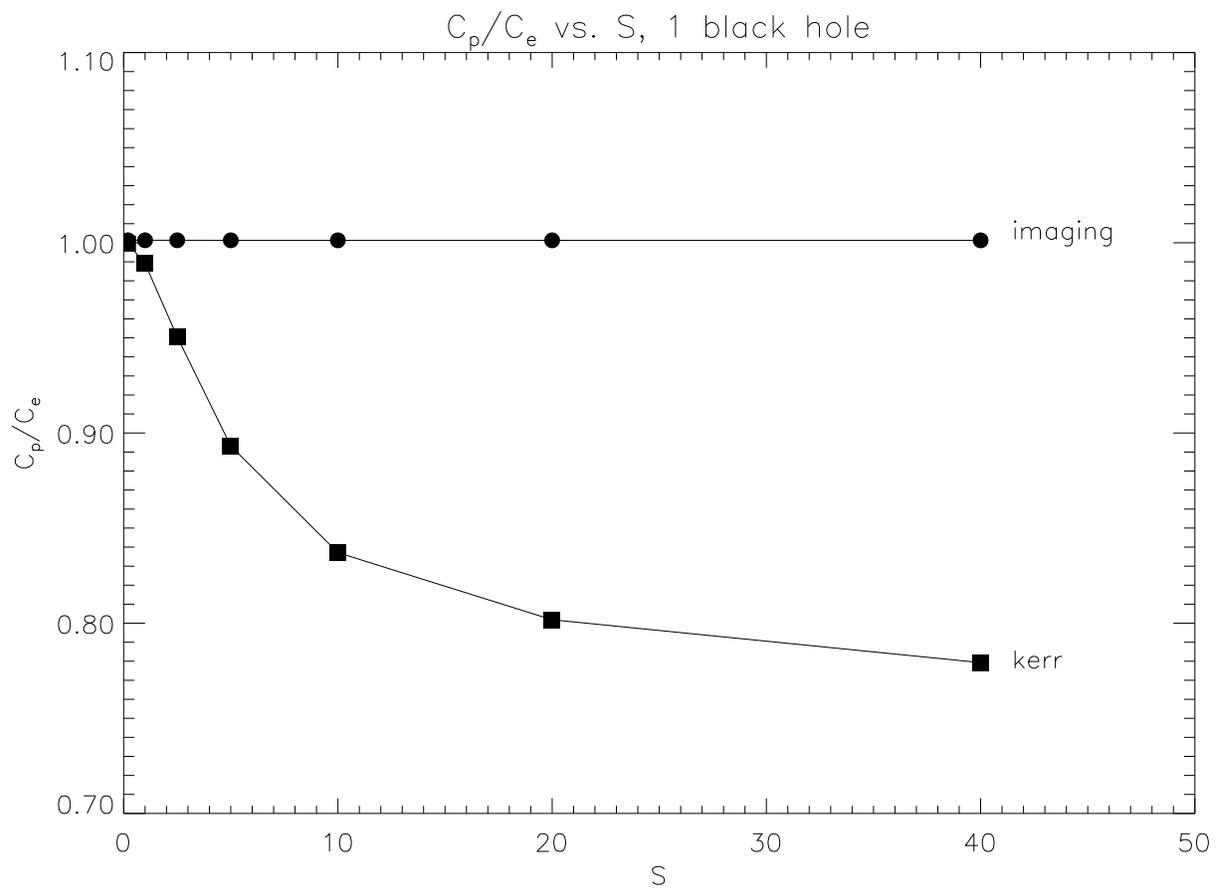,width=17cm}
\caption[]{This plot shows polar circumference/equatorial
circumference, $C_p/C_e$, for a black hole calculated by the imaging
method and a Kerr black hole with the same overall mass and angular
momentum, as a function of S. 
\label{fig:4}}
\end{center}
\end{figure}

In the conformal-imaging approach, the location of an apparent horizon
for a single rotating black hole coincides with the throat
\cite{Cook3}. Thus, in this case it is possible to investigate the
properties of the apparent horizon. We have computed the area of the
horizon $A_{ah}$, the mass, $M_{ah}=\sqrt{A_{ah}/16\pi}$ and the polar
and equatorial circumference, $C_p$ and $C_e$. The values of $M_{ah}$
agree within three digits with the corresponding values in
\cite{Cook3} for $S=10$.  Figure \ref{fig:3} shows calculated
$M_{tot}$ and $M_{ah}$ as a function of $S$ for a rotating black hole
with $M=2$.  Also shown is the irreducible mass $M_{irr}$ calculated
for a Kerr black hole with the corresponding $S$ and $M_{tot}$. Figure
\ref{fig:4} shows the calculated ratio of the polar and equatorial
circumferences $C_p/C_e$ as a function of $S$ and compares it with the
corresponding ratio for a Kerr black hole \cite{Smarr}.

Figure \ref{fig:3} shows that for the same total mass of the
configuration $M_{tot}$ and the same angular momentum $S$, the mass of
the apparent horizon in the numerical data, $M_{ah}$, is less than the
irreducible mass of the black hole. This is expected since in the
conformal approach there is an additional energy/mass related to a
junk field outside of the black hole. Figure \ref{fig:4} shows that
$C_p/C_e$ is close to unity for conformal imaging solutions whereas
for Kerr black hole it is less than one. This means that the shape of
the apparent horizon is substantially closer to spherical symmetry in
the conformal-imaging solutions than for a Kerr black hole.

We have also compared the scalar invariants (17) of our numerical solutions
with the corresponding analytical values. The difference between the
values of these invariants of the numerical and and analytical
solutions characterize the junk gravitational field. We have chosen to
compare the value of the invariant in the points that are the same
physical distance from the apparent horizon along the equatorial axis
of the black hole, because this gives us a correspondence between the
Boyer-Lindquist radial coordinate $r$ of the Kerr metric and the
isotropic radial coordinate $x$ of the numerical solution. In the
numerical solution the apparent horizon coincides with the throat,
and in the Kerr solution the apparent horizon coincides with the event
horizon. To find the Boyer-Lindquist radial coordinate $r$ that
corresponds to a given isotropic coordinate $x$ we find the physical
distance:
\begin{equation}
\label{eq:22}
\Delta R_{ph}(x) = \int_{R}^x \Psi^2 dx~,
\end{equation}
where $x$ is the distance to the center of the black hole throat in
the background flat space and $R$ is the throat radius. The
integration is performed numerically using cubic spline quadrature. We
normalize this distance:
\begin{equation}
\label{eq:23}
\Delta l_{ph}(x) = \Delta R_{ph} \frac{R}{M_{tot}}
\end{equation}
and we write down the following implicit equation:
\begin{equation}
\label{eq:24}
  \Delta l_{ph}(x) M_{tot} = \int_{M_{tot} + \sqrt{M_{tot}^2 -
a^2}}^r \sqrt{g_{rr}} dr
\end{equation}
and solve it with respect to $r$ using Van Wijngaarden-Dekker-Brent's
method. $M_{tot}$ is the total mass of the numerical black hole with
mass parameter $M=2R$, as well as the total mass of the Kerr black
hole that we are comparing to. Finally we compute the value of the
invariants for the Kerr black hole (which are known analytically) at
$r(x)$.

\begin{figure}
\begin{center}
\epsfig{file=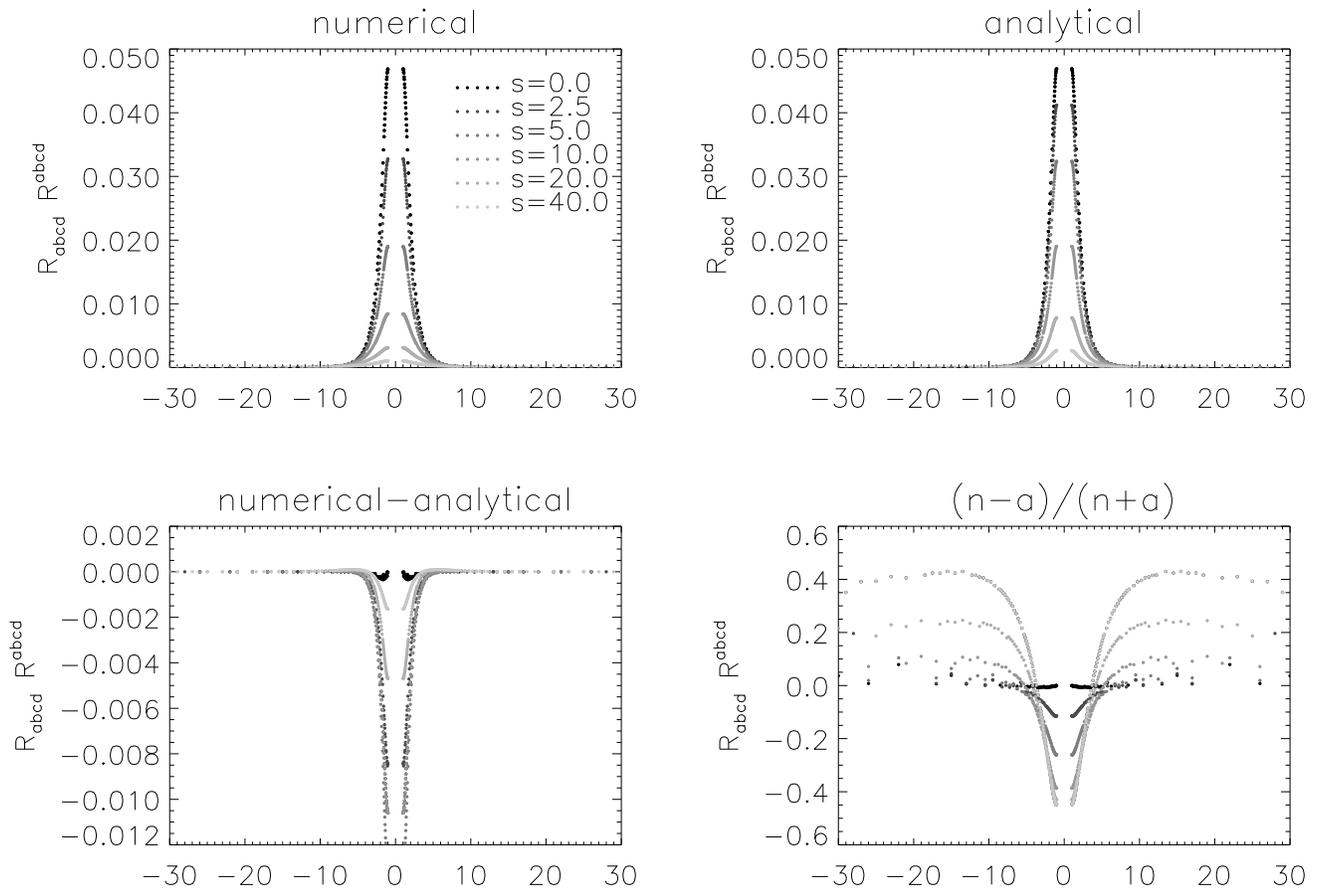}
\caption[]{$I_1$ (eq. \ref{eq:17}), as a function of
isotropic coordinate y (which coincides with the equatorial axis) for 
rotating numerical and Kerr black holes.
\label{fig:5}}
\end{center}
\end{figure}

\begin{figure}
\begin{center}
\epsfig{file=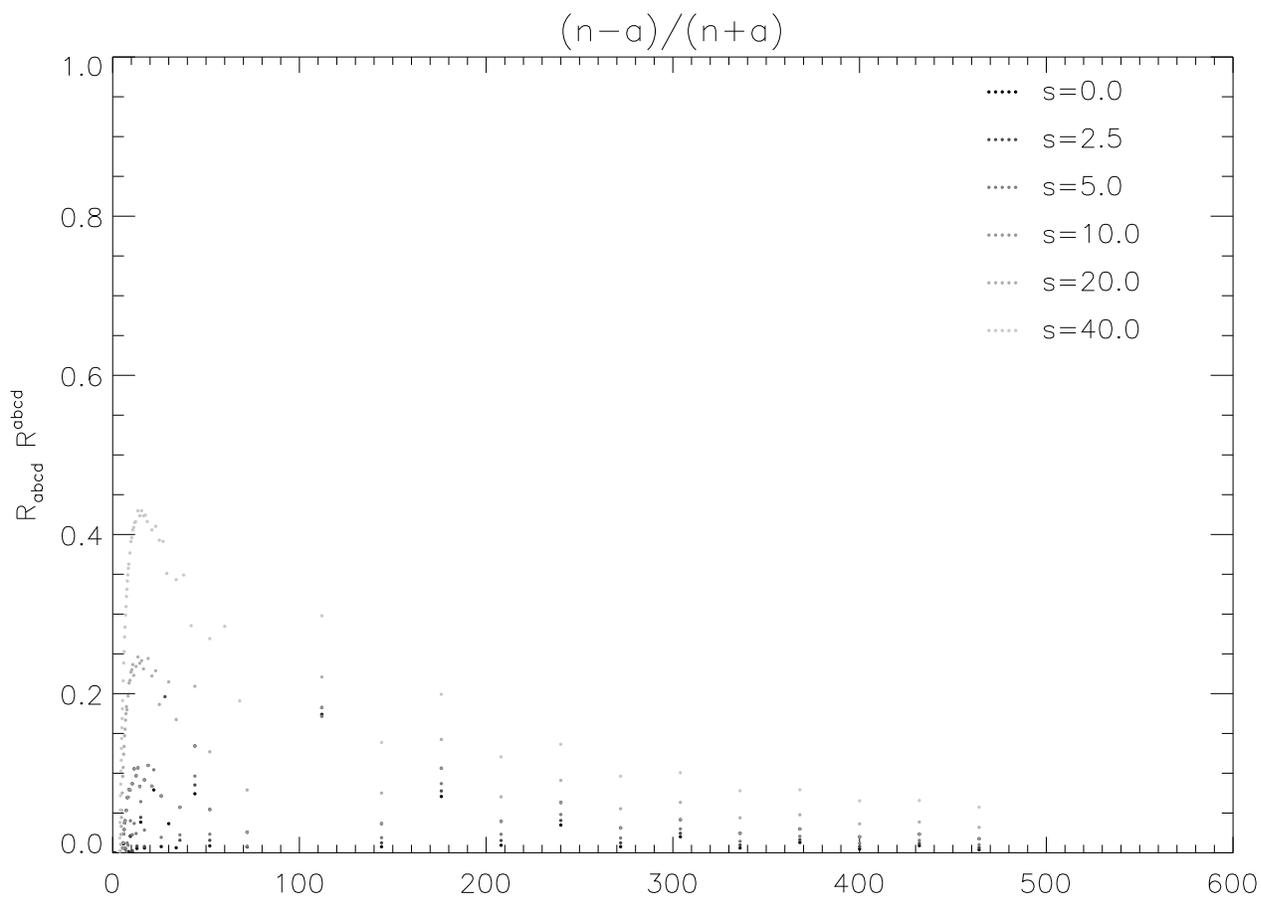}
\caption[]{Relative comparison of $I_1$ (eq. \ref{eq:17}) 
at large distances for rotating holes.
\label{fig:6}}
\end{center}
\end{figure}

Figure (\ref{fig:5}) shows that increasing value of $S$ leads to the
decrease of $I_1$ at the black hole horizon for Kerr black holes.
This is a well understood effect: from (\ref{fig:3}) we know that the
total mass of a black hole and its apparent horizon grow with $S$ and
we know that in the Kerr solution tidal forces at the event horizon of
a Kerr black hole decrease with the increase of the mass. We can see
that the same tendency holds for conformal black holes (Figure
\ref{fig:5} a,b). Figure \ref{fig:5} c shows the difference in $I_1$
for the conformal and Kerr solutions. At the boundary of the black
hole the absolute value of the difference initially decreases with
increasing $S$. However, for large $S$ it begins to decrease due to
the decrease of tidal forces. Figure~\ref{fig:5}d shows the relative
difference in $I_1$. The absolute value of the relative difference at
the black hole boundary increases with increasing $S$
monotonically. Far away from the black hole, the absolute value of the
difference is also monotonically increases with $S$, and for large
values of $S$ it is significantly larger than that at the black hole
boundary. Figure \ref{fig:3} thus shows that there exists a junk
gravitational field in the conformal solutions, and that the relative
amount of this junk field increases outside of the black hole with
increasing $S$. Figure \ref{fig:6} shows the relative difference of
$I_1$ at large distances. The relative difference is remains large
even at very large $r\simeq 500$.

\subsection{\bf \normalfont \normalsize 
\it Moving single black hole} 

\begin{figure}
\begin{center}
\epsfig{file=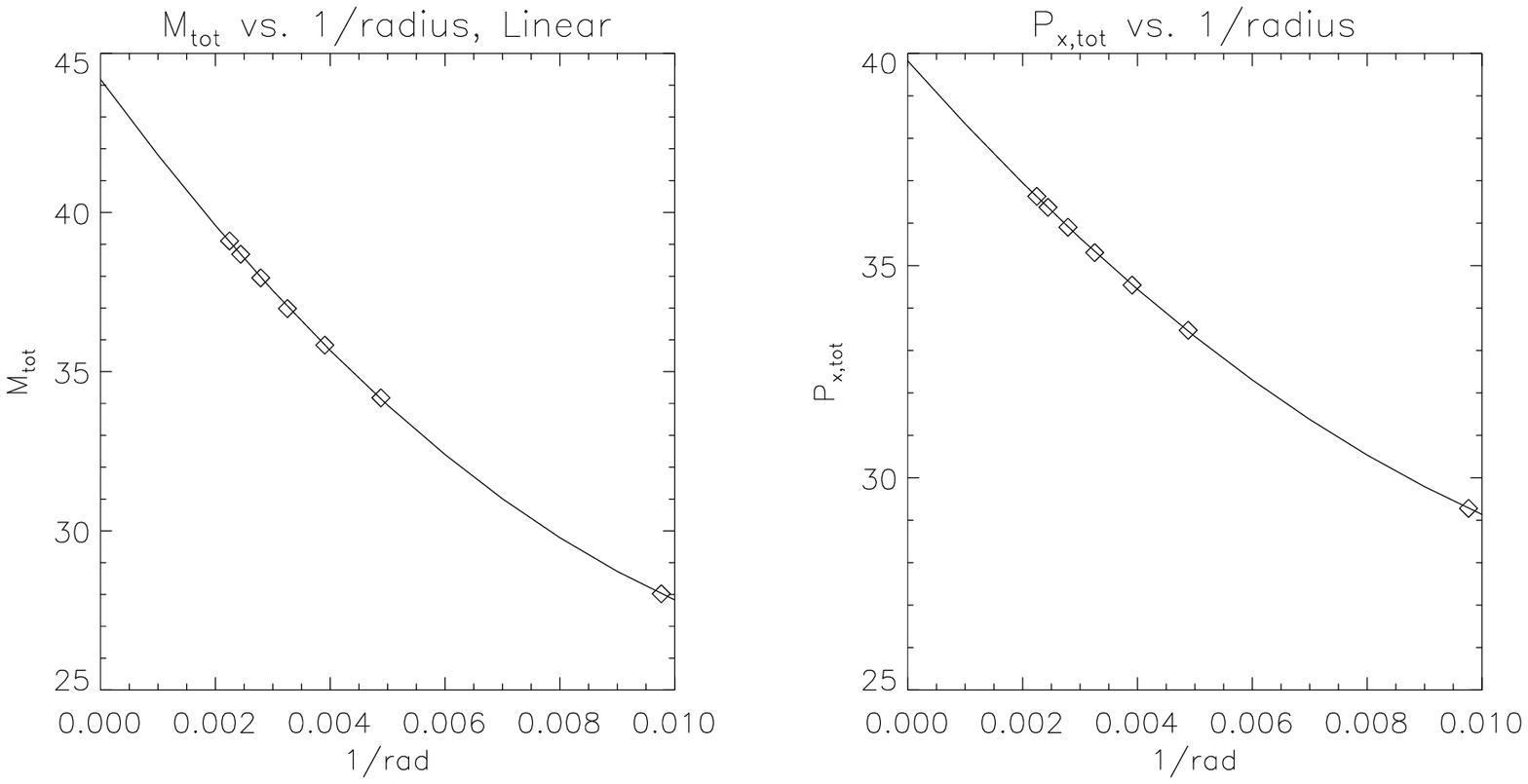,width=17cm}
\caption[]{Asymptotic total mass, $M_{tot}$, and total linear
momentum, $P_{tot}$, for case M=2 P=40.
The diamonds are the numerical values of $M_{tot}$,$P_{tot}$ found
from integrating over a sphere at radius $r$. The line is a 
least square fit of the data to a second order polynomial.
\label{fig:7}}
\end{center}
\end{figure}

Now consider a single black hole with a non-zero linear momentum ${\bf
P}\neq 0$ and zero angular momentum ${\bf S}= 0$. Similar to rotating
black holes, we compute numerical conformal solutions with $M=2$ and
various $P$, calculate $M_{tot}$ and $P_{tot}$ for these solutions by
integrating (12) and (13) over spherical surfaces of different radii
$r$, and then extrapolate these values to infinity.  Figure
\ref{fig:7} illustrates the extrapolation procedure for the case $M=2$
and $P=40$. In the case of a boosted black hole, both $M_{tot}$ and
$P_{tot}$ vary noticeably with $r$. The extrapolated value of
$P_{tot}(\infty) \simeq 40\simeq P$, as it should be.  We compare our
numerical solutions with known analytical solutions for a boosted
Schwarzschild black hole with the same $M_{tot}(\infty)$ and
$P_{tot}(\infty)$.

The spatial metric of a boosted Schwarzschild black hole can be found
by performing a Lorenz transformation
\begin{eqnarray}
\label{eq:25}
\bar t &=& \gamma (t + v x)~,\\
\bar x &=& \gamma (x + v t) \\
\end{eqnarray}
of a Schwarzschild metric, where $\gamma = (1 - v^2)^{1/2}$.  This
gives the following spacetime metric
\begin{equation}
\label{eq:26}
ds^2 = -\alpha \sqrt{\frac{1}{\gamma^2 \left(1 - v^2
\frac{\alpha^2}{\Psi^4} \right)}}
d\bar{t}^2 
+ \gamma^2 \left(\Psi^4 - \alpha^2 v^2\right)\left(d\bar{x} -
v \frac{1 - \frac{\alpha^2}{\Psi^4}}{1 - v^2
\frac{\alpha^2}{\Psi^4}} d\bar{t}\right)^2 \Psi^4 \left(d\bar{y}^2 +
d\bar{z}^2\right)
\end{equation}
with
\begin{equation}
\label{eq:27}
\Psi =\left(1 + \frac{M}{2 R}\right)~, \qquad  \alpha =
\frac{1 - \frac{M}{2 R}}{1 + \frac{M}{2 R}} ~,\qquad
R = \sqrt{x^2 + y^2 + z^2}~.
\end{equation}
The analytical solution (\ref{eq:26}) is described by the two
parameters, the mass of the Schwarzschild black hole $M$ and the
velocity of the hole $v$ as measured by an observer at infinity.  The
total mass and velocity of a boosted Schwarzschild black hole are
\begin{equation}
\label{eq:28}
M_{tot} = \sqrt{M^2 + P^2} ~,\qquad v = \sqrt{\frac{P^2}{M^2 + P^2}} =
\frac{P}{M_{tot}} ~.
\end{equation}

At the horizon, the metric is exactly the same as for a Schwarzschild
black hole (this can be seen from the fact that $\alpha = 0$ at the
horizon). Thus, the horizon properties are the same as for a
Schwarzschild black hole. Also, since we have only performed a
coordinate transformation, $I_1$ for a boosted Schwarzschild black
hole is the same as $I_1$ for a Schwarzschild black hole.

\begin{figure}
\begin{center}
\epsfig{file=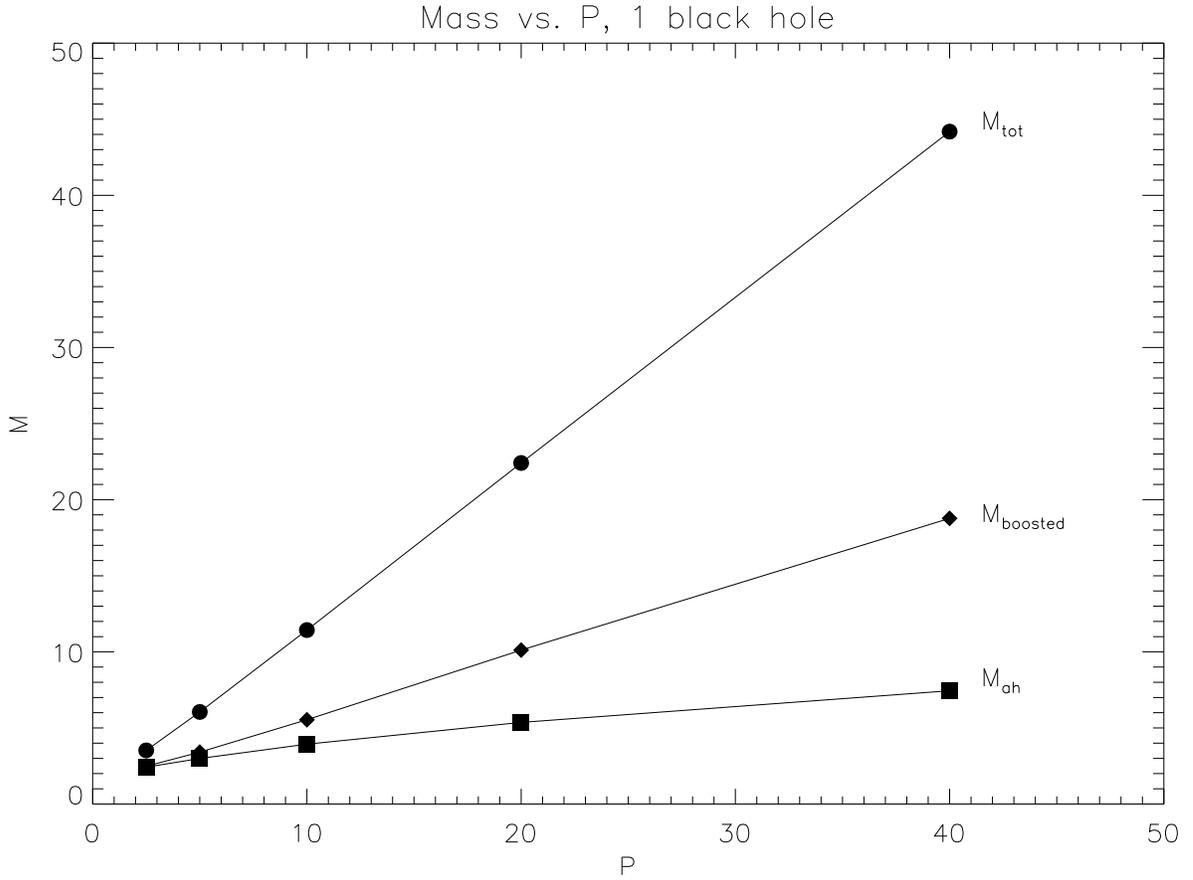,width=17cm}
\caption[]{Total mass, $M_{tot}$, apparent horizon mass, $M_{ah}$, and
boosted mass, $M_{boosted}$, as a function of P for a 
boosted black hole. $M_{tot}$ and $M_{ah}$ are calculated
numerically, and $M_{boosted} = \sqrt{M_{tot}^2 - P^2}$ is the horizon
mass of a boosted Schwarzschild black hole with the same overall 
mass and linear momentum as the numerical black hole.
\label{fig:8}}
\end{center}
\end{figure}

In the conformal-imaging approach, the location of an apparent horizon
for a moving black hole coincides with the throat \cite{Cook3} and for
a Schwarschild black hole the apparent horizon coincides with the
event horizon. We have computed the area of the horizon $A_{ah}$, the
mass, $M_{ah}=\sqrt{A_{ah}/16\pi}$ and the polar and equatorial
circumference, $C_p$ and $C_e$.  Figure \ref{fig:8} shows calculated
$M_{tot}$ and $M_{ah}$ as a function of $P$ for a moving black hole
with $M=2$. Also shown is the analytical horizon mass $M_{boosted} =
\sqrt{M_{tot}^2 - P^2}$.  Figure \ref{fig:8} shows that for the same
total mass of the configuration $M_{tot}$ and the same linear momentum
$P$, the mass of the calculated apparent horizon $M_{ah}$ is less than
the analytical one. This is expected since in the conformal approach
there is an additional energy related to a junk field outside of the
black hole. The polar and equatorial circumferences for a moving black
hole are practically the same, i.e. the symmetry of the horizon is
close to spherical symmetry.

\begin{figure}
\begin{center}
\epsfig{file=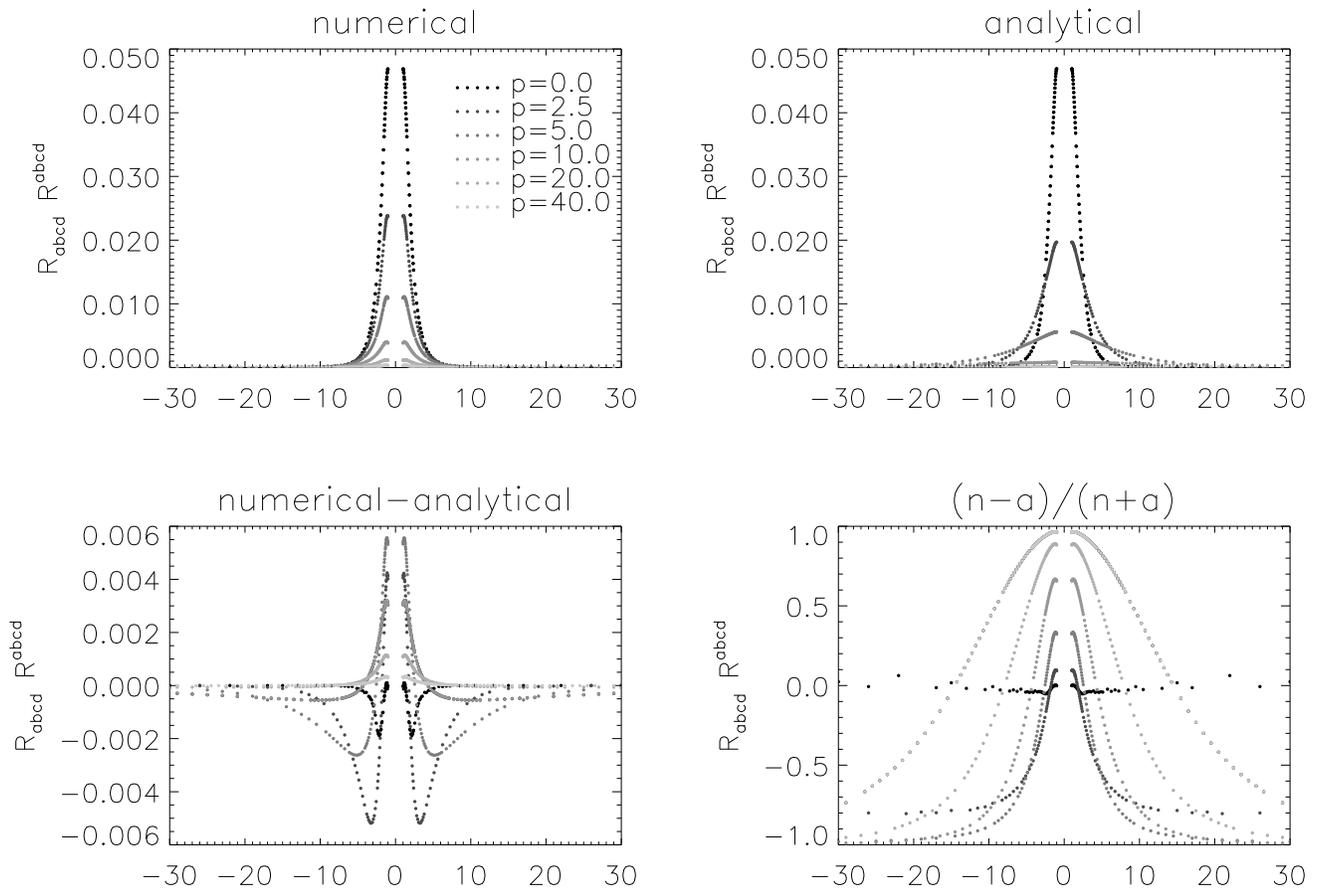}
\caption[]{$I_1$ (eq. \ref{eq:17}), as a function of
isotropic coordinate x (which coincides with the equatorial axis) for 
moving numerical and boosted Schwarzschild black holes.
\label{fig:9}}
\end{center}
\end{figure}

\begin{figure}
\begin{center}
\epsfig{file=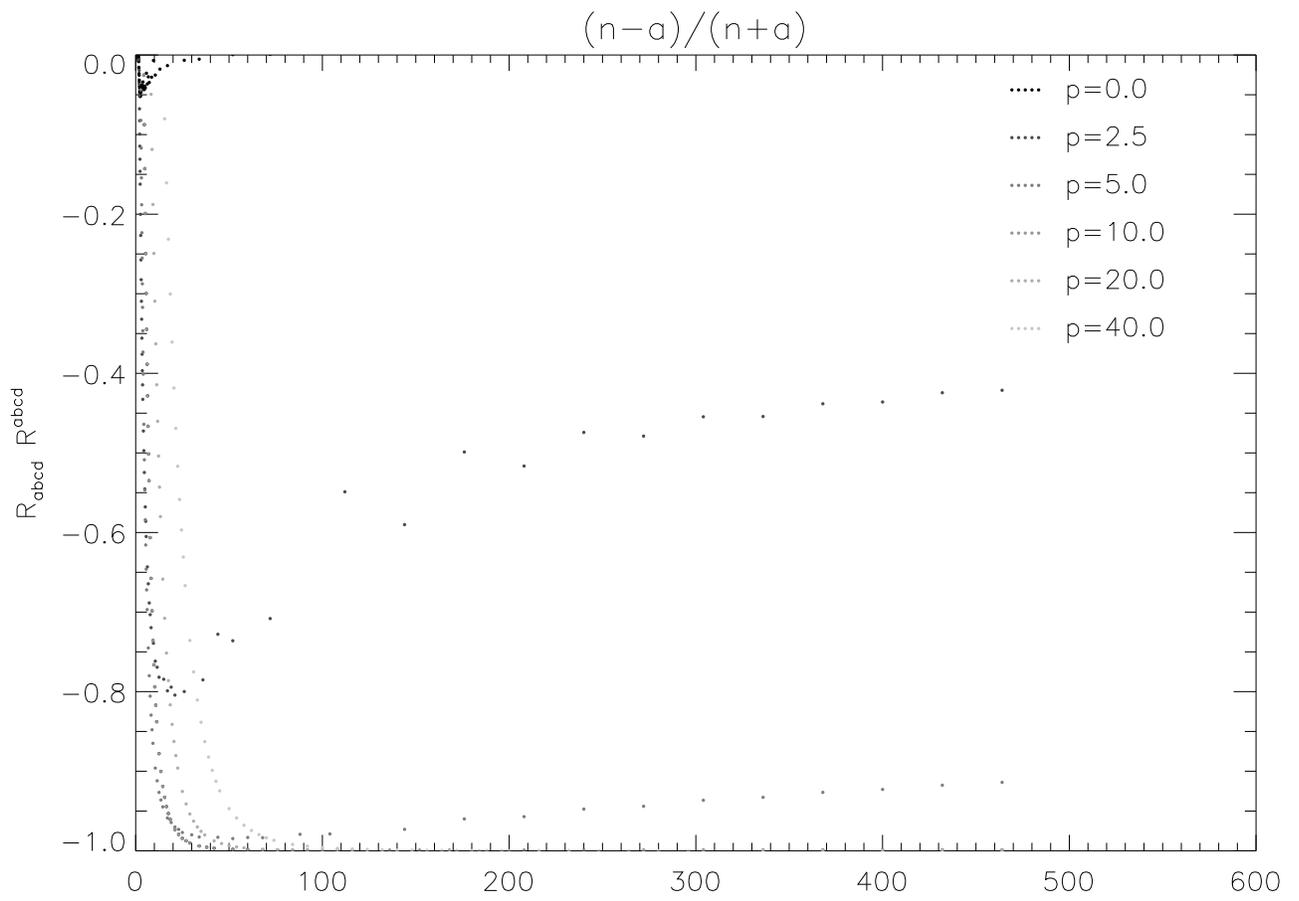}
\caption[]{Relative comparison of $I_1$ (eq. \ref{eq:17}) at large 
distances for boosted black holes.
\label{fig:10}}
\end{center}
\end{figure}

We have compared the scalar invariants (17) of our numerical solutions
with the corresponding analytical values.  We have chosen to compare
the value of the invariant in the points that are the same physical
distance from the apparent horizon along the axis of motion (the
x-axis). This physical distance can be expressed as a function of the
isotropic coordinate distance $x$ from the black hole center along the
axis of motion to the point where we make the comparison. Figure
\ref{fig:9} compares $I_1$ as a function of $x$ for various values of
$P$. We see that the increase of the total mass with increasing $P$
leads to a decrease in the value of $I_1$ at the black hole horizon
(Figure \ref{fig:9} a,b). Figure \ref{fig:9}c shows the difference in
$I_1$ for numerical and analytical solutions.  At the black hole
boundary, the absolute value of the difference initially decreases
with increasing $P$. However for large $P$ it begins to decrease due
to the decrease of tidal forces. Figure \ref{fig:9}d shows the
relative difference in $I_1$. The absolute value of the relative
difference at the black hole boundary first increases with increasing
$P$ but than starts do decrease. Far away from the black hole,
values of $I_1$ for boosted black holes are much larger than $I_1$ for
conformal black hole solutions (Figure \ref{fig:10}).  Similar to the
rotating black hole case, Figure \ref{fig:9} show that conformal
solutions contain a junk gravitational field. The amount of the junk
field increases outside of the black hole with increasing $P$.

\section{Binary black holes}
\label{sec:5}
\setlength{\parindent}{0em} \setlength{\parskip}{1.5ex plus 0.5ex}
There is an infinite number of possibilities for choosing the
parameters that characterize the initial data for two black holes. The
problem can be parameterized by
\begin{equation}
\label{eq:29}
(r_2, r_1, d = \frac{\|C_1^i - C_2^i\|}{2 r_1}, P_1^i, S_1^i, P_2^i, S_2^i)
\end{equation}
where $r_\delta = \frac{M_\delta}{2}$ and $M_\delta$ is the mass
parameter of the $\delta$'th black hole. $C_1$ and $C_2$ are the
coordinates of the center of the throats and punctures respectively.
We chose to call the distance $d=\frac{\|C_1^i - C_2^i\|}{2 r_1}$ the
separation of the hole. Since the holes are placed at equal distance
from the center of the computational domain, $d$ is the coordinate
distance from the center of each hole to the center of the domain. 
The rest of the parameters are the linear/angular momentum parameters.

\begin{figure}[ht]
\begin{center}
\begin{minipage}[b]{6.2cm}
A)\epsfig{file=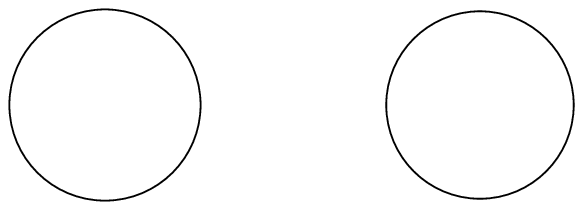,width=17cm,width=6.0cm}
\end{minipage}
\begin{minipage}[b]{6.2cm}
B)\epsfig{file=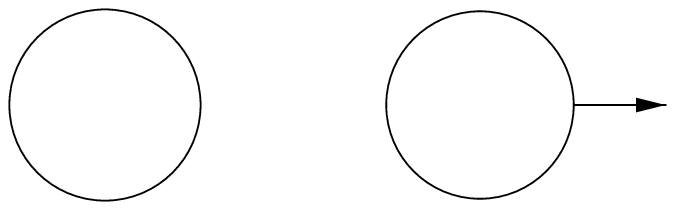,width=17cm,width=6.0cm}
\end{minipage}
\begin{minipage}[b]{6.2cm}
C)\epsfig{file=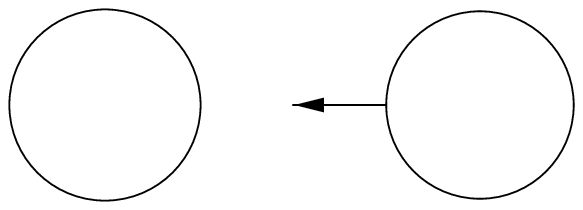,width=17cm,width=6.0cm}
\end{minipage}
\begin{minipage}[b]{6.2cm}
D)\epsfig{file=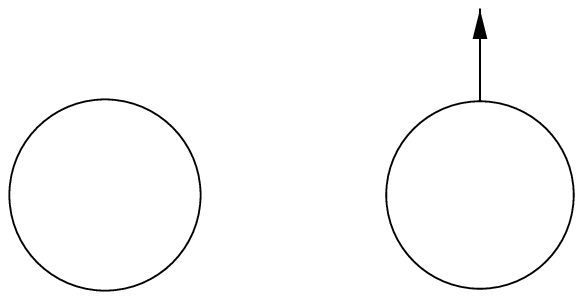,width=17cm,width=6.0cm}
\end{minipage}
\begin{minipage}[b]{6.2cm}
E)\epsfig{file=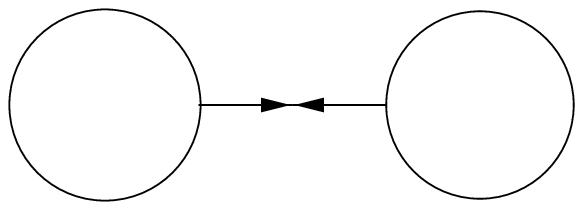,width=17cm,width=6.0cm}
\end{minipage}
\begin{minipage}[b]{6.2cm}
F)\epsfig{file=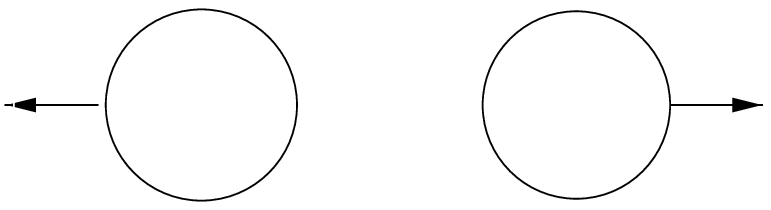,width=17cm,width=6.0cm}
\end{minipage}
\begin{minipage}[b]{6.2cm}
G)\epsfig{file=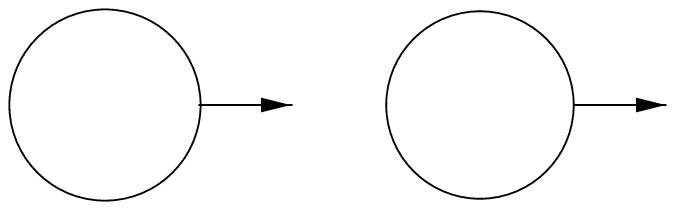,width=17cm,width=6.0cm}
\end{minipage}
\begin{minipage}[b]{6.2cm}
H)\epsfig{file=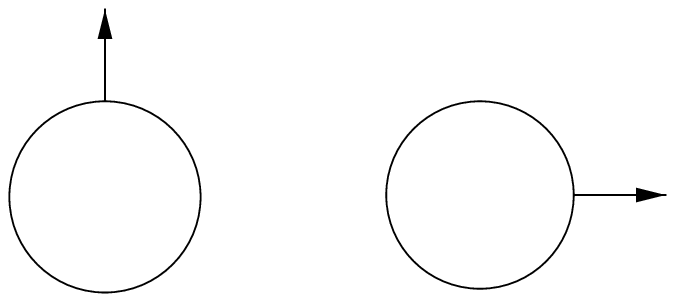,width=17cm,width=6.0cm}
\end{minipage}
\begin{minipage}[b]{6.2cm}
I)\epsfig{file=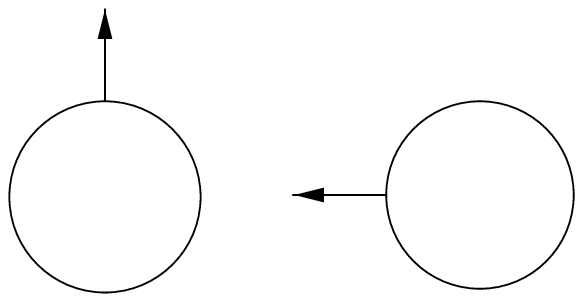,width=17cm,width=6.0cm}
\end{minipage}
\begin{minipage}[b]{6.2cm}
J)\epsfig{file=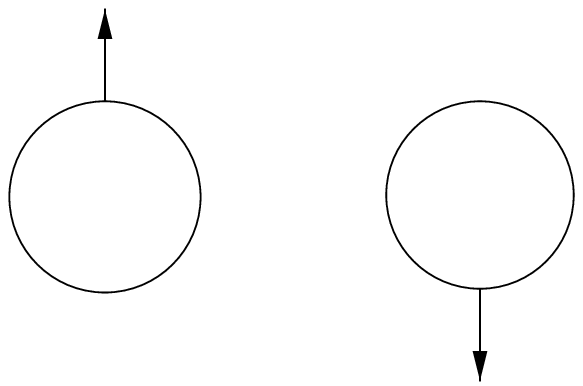,width=17cm,width=6.0cm}
\end{minipage}
\caption[]{Schematic representation of 2 black hole configurations
\label{fig:11}}
\end{center}
\end{figure}

We restrict ourselves to some specific case. We consider
configurations with equal mass black holes, $M_1=M_2=M=2$, i.e.
$R_1=R_2=1$. The line connecting the centers of the holes coincides
with the x-axis. We only investigate configurations where the holes
have either linear momenta or angular momenta, but not both. The
configurations considered are shown schematically on Figure
\ref{fig:11}. Arrows on this figure show the direction of either
linear or angular momenta. In all cases when momenta of both black
holes are not zero, they have equal absolute values except when
explicitly stated otherwise.

First we mention the following general property of our initial data.
Let the linear momentum parameters of the problem are ${\bf P}_1$ and
${\bf P}_2$, and the angular momentum parameters for both holes are
zero. Then the metric that is a solution to the constraint equations,
(\ref{eq:4}) and (\ref{eq:5}) with these parameters is the same as the
metric of that is the solution to the constraint equations where the
linear momentum parameters are $-{\bf P}_1$ and $-{\bf P}_2$, i.e. the
problem where the sign of all components of the linear momentum
parameters has been reversed.  The proof can be found in appendix A.
It is also possible to prove that a solution where the parameters
${\bf P}_1, {\bf P}_2 = 0$ and ${\bf S}_1$ and ${\bf S}_2$ has some
non-zero values will have the same metric as a solution with
parameters $-{\bf S}_1$ and $-{\bf S}_2$. Therefore, configurations E
and F (figure \ref{fig:11}), will have the same total mass
$M_{tot}$. The same holds for configurations B and C.

For configurations with linear momenta, we will compare numerical
results for $M_{tot}$ with the predictions of special theory of
relativity,
\begin{equation}
\label{eq:30}
M_{tot}^{STR} = \sqrt{M_1^2+P_1^2} + \sqrt{M_2^2 + P_2^2}       ~.
\end{equation}
Equation (\ref{eq:30}) ignores gravitational interaction between the
holes. We also can compare $M_{tot}$ of a two black hole
configuration $(M,P_1,P_2)$ with the sum $M_{tot}^*$ of numerical
$M_{tot}$ of individual black holes with parameters $(M,P_1)$ and
$(M,P_2)$,
\begin{equation}
\label{eq:31}
M_{tot}^* = M_{tot}(M,P_1) +
M_{tot}(M,P_2) ~. 
\end{equation}
Again, $M_{tot}^*$ ignores mutual gravitational interaction of the
black holes.

For configurations with angular momenta, we can compare numerical
$M_{tot}$ with 
\begin{equation}
\label{eq:33}
M_{tot}^*  = M_{tot}(M,S_1) +
M_{tot}(M,S_2) ~. 
\end{equation}
similar to the linear momentum case.

Now we discuss configurations without linear or angular momentum and
configurations with linear momentum $P=5$ and with varying separation
$d$ between the holes. For these configurations, we found that the
difference between $M_{tot}$ extrapolated to $r=\infty$ and $M_{tot}$
evaluated at $r=512$ is of the order of 1\%.  In what follows, we
present values of $M_{tot}$ evaluated at $r=512$.

\subsection{\bf \normalfont \normalsize 
\it Two black holes with zero momenta} 

\begin{figure}
\begin{center}
\epsfig{file=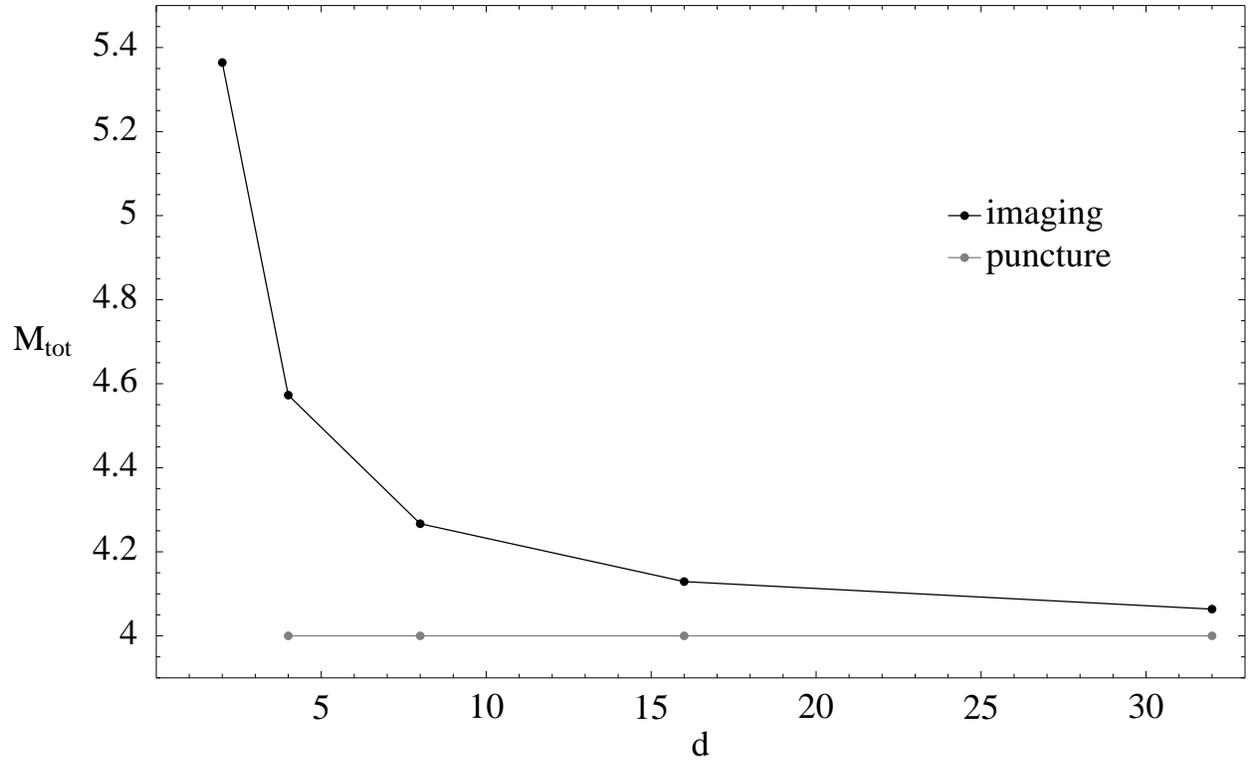,width=17cm}
\caption[]{Total mass, $M_{tot}$ as a function of separation, $d$, 
for two black holes with configuration A (figure \ref{fig:11})
\label{fig:12}}
\end{center}
\end{figure}

Figure \ref{fig:12} shows $M_{tot}$ as a function of $d$ for Case A,
the configuration of two black holes without linear or angular
momentum. The exact solution for this case can be found with the
semi-analytic method described in \cite{Misner}. As one can see, the
total mass converges to 4.0 when the distance between the holes
increases. This is expected, because the gravitational attraction
between the holes (which also contributes to the total mass) becomes
negligible as the distance between the holes gets larger, and the
holes can be regarded as point particles. However, the gravitational
attraction should decrease the total mass, but we see that $M_{tot}$
increases or is constant with decreasing $d$. This behavior clearly
shows the presence of an extra gravitational field that artificially
cancels the effects of gravity on the total mass (in the puncture
data) or adds even more energy that the amount required to cancel out
the effect on the total mass (in the imaging data). The puncture
solutions contain less junk that the imaging solutions.

\subsection{\bf \normalfont \normalsize 
\it Two black holes with linear momenta}

\begin{figure}
\begin{center}
\epsfig{file=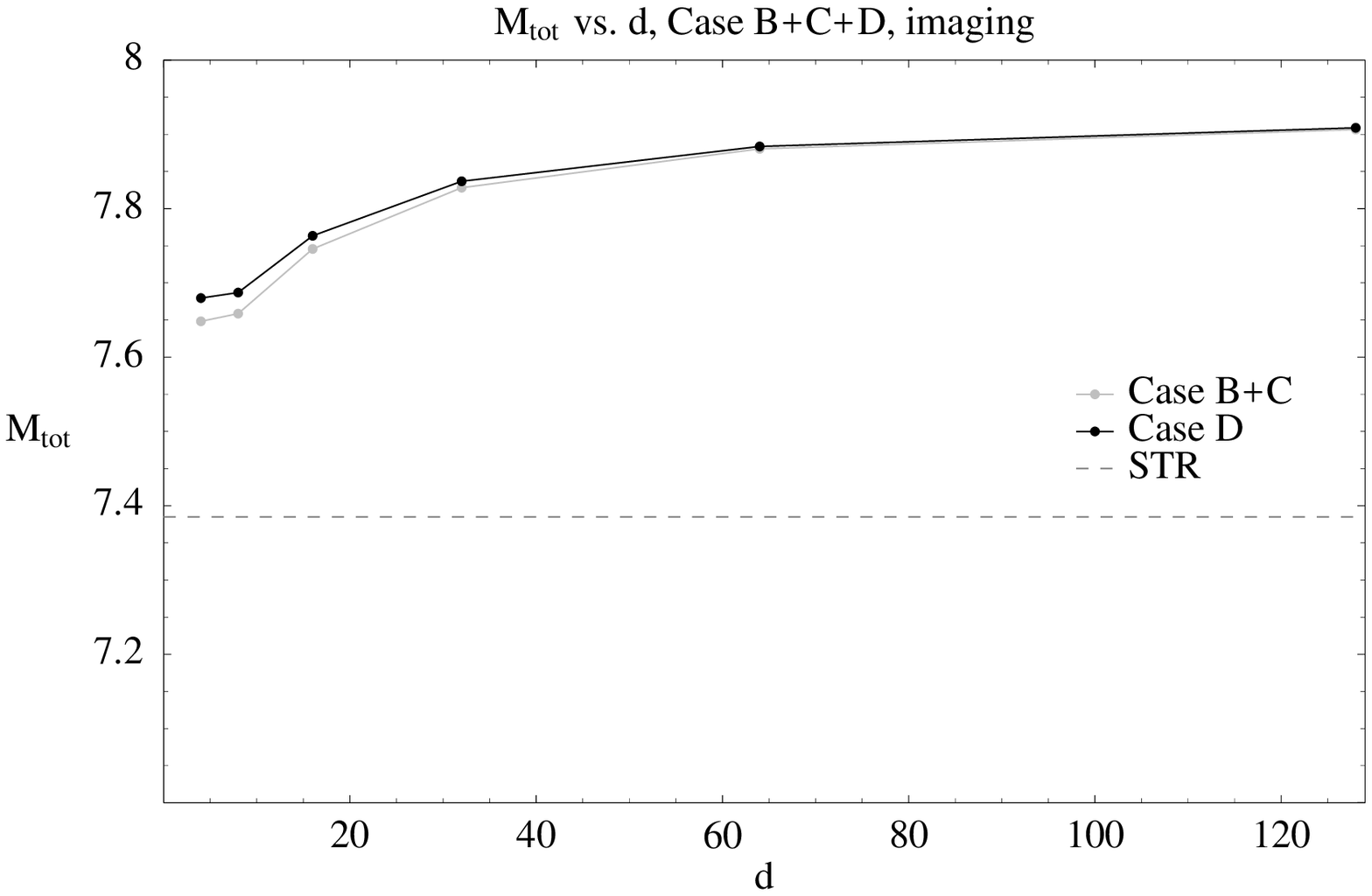,width=17cm}
\caption[]{Total mass, $M_{tot}$, as a function of separation, $d$, 
for two black holes with configurations B, C, D (figure \ref{fig:11}) with
linear momentum, imaging method
\label{fig:13A}}
\end{center}
\end{figure}

\begin{figure}
\begin{center}
\epsfig{file=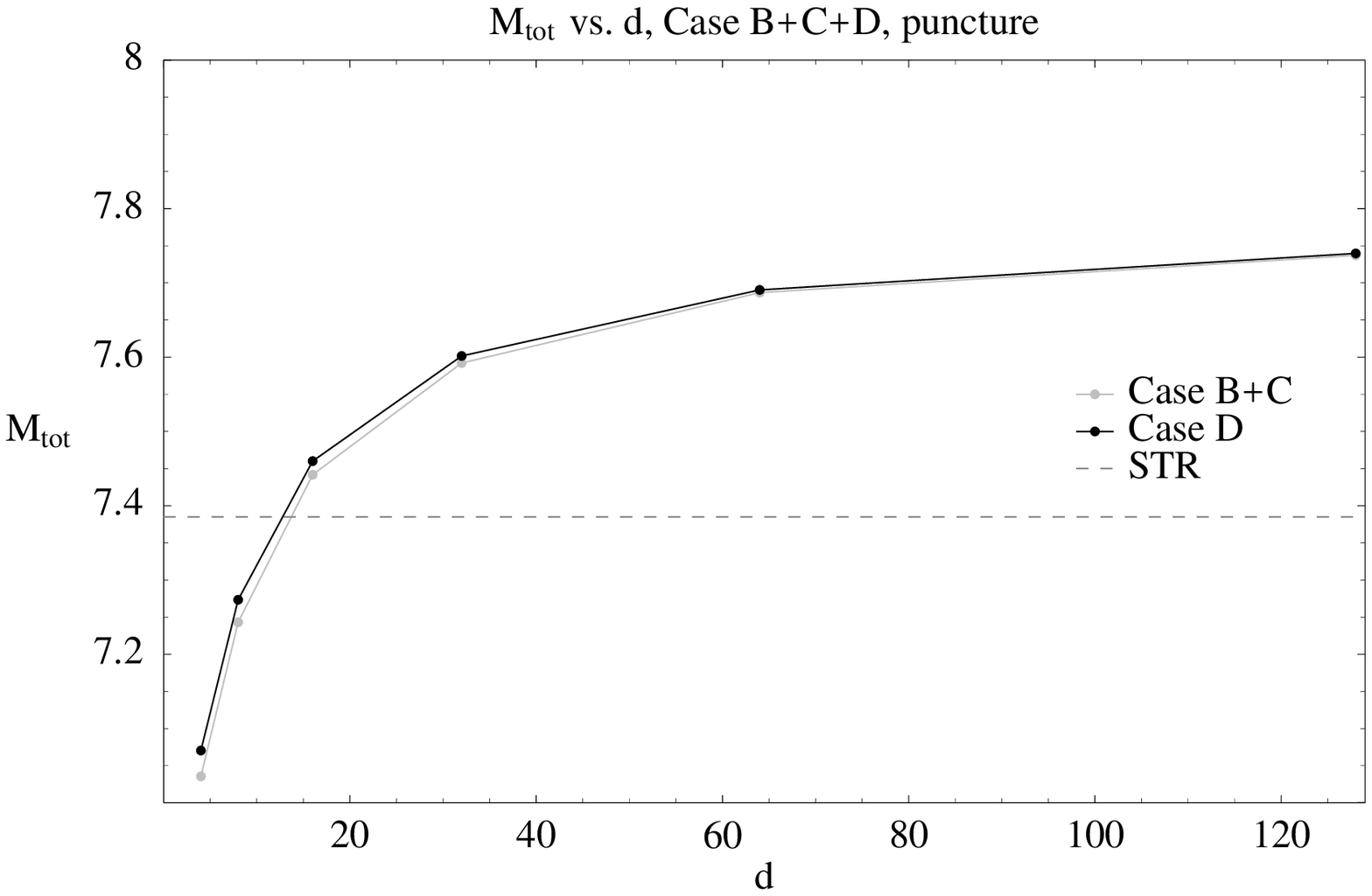,width=17cm}
\caption[]{Total mass, $M_{tot}$, as a function of separation, $d$, 
for two black holes with configurations B, C, D (figure \ref{fig:11}) with
linear momentum, puncture method
\label{fig:13B}}
\end{center}
\end{figure}

Figures \ref{fig:13A} and \ref{fig:13B} show $M_{tot}$ for cases B, C
and D which contain one black hole with linear momentum $P=5$ another
with $P=0$. The figures show that, in both conformal-imaging and
puncture methods, $M_{tot}$ is practically independent of orientation
of $\bf P$ with respect to the line passing through the black hole
centers. At large separations, $M_{tot}$ for both methods does not
tend to the special relativity limit $M_{tot}^{STR}$. Instead $M_{tot}$
tends to $M_{tot}^*$ (Equation (28)). This shows that even at larger
separations, the solutions contain a junk gravitational field. This
junk field is associated with the junk fields of individual solution
for black hole with $P=5$. From the comparison at small separations we
see that puncture solutions contain less junk than conformal imaging
solutions.

\begin{figure}
\begin{center}
\epsfig{file=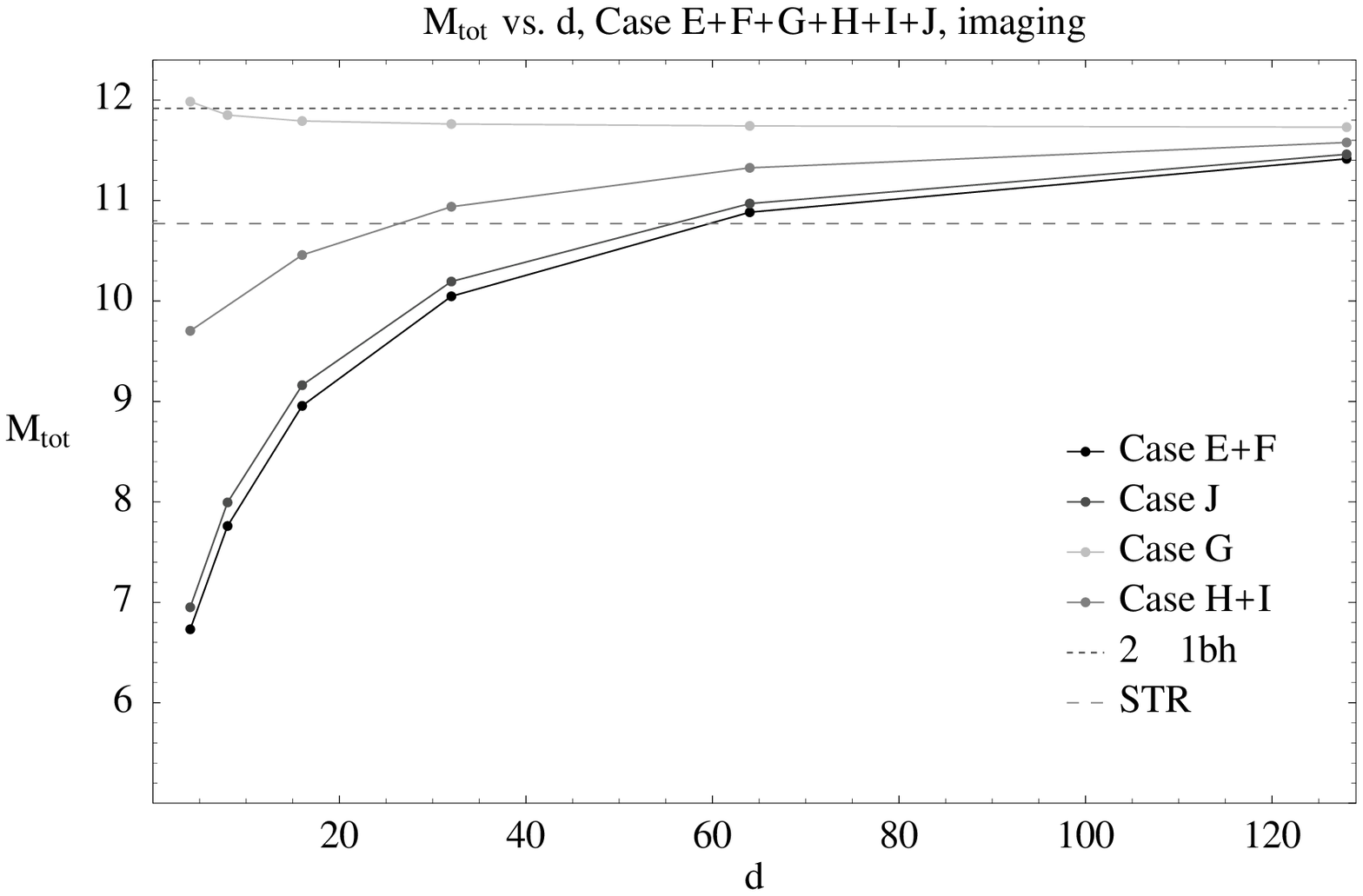,width=17cm}
\caption[]{Total mass, $M_{tot}$, as a function of separation, $d$, 
for two black holes with configurations E, F, G, H, I, J 
(figure \ref{fig:11}) with linear momentum, imaging method
\label{fig:14A}}
\end{center}
\end{figure}

\begin{figure}
\begin{center}
\epsfig{file=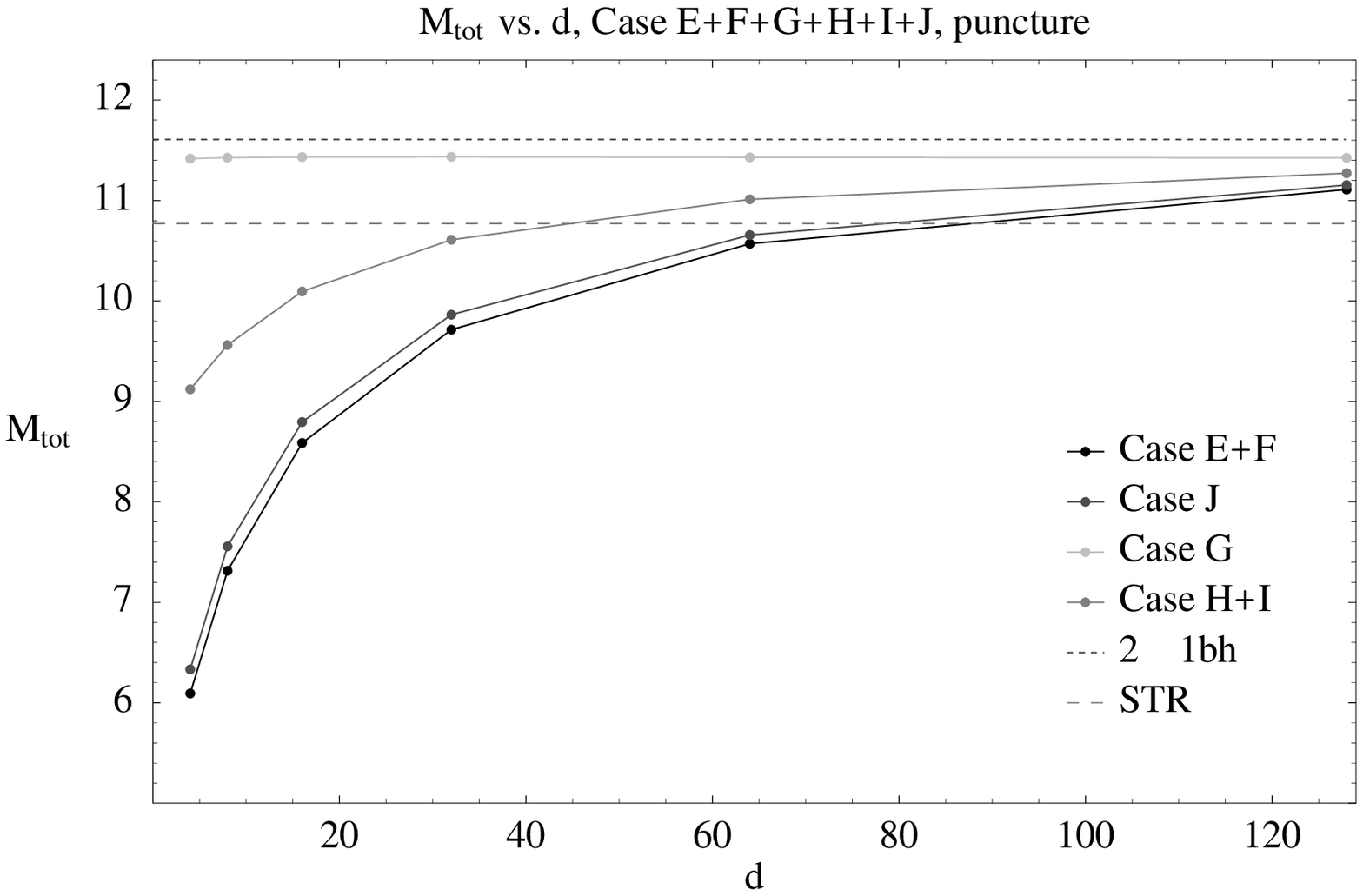,width=17cm}
\caption[]{Total mass, $M_{tot}$, as a function of separation, $d$, 
for two black holes with configurations E, F, G, H, I, J 
(figure \ref{fig:11}) with linear momentum, puncture method
\label{fig:14B}}
\end{center}
\end{figure}

Figures \ref{fig:14A} and \ref{fig:14B} show $M_{tot}$ for cases where
both black holes have linear momentum $P=5$ but with different
relative orientations.  We see that $M_{tot}$ for these cases depend
on relative orientation of $\bf P$.  When ${\bf P}_1$ and ${\bf P}_2$
are antiparallel, $M_{tot}$ is practically the same, independent of
orientation of $\bf P$ (compare cases E, F and J). These
configurations also have minimal masses compared to other cases at the
same separations. Configuration G with parallel momenta gives the
largest $M_{tot}$ which is almost independent of $d$. Configurations H
and I with perpendicular momenta are intermediate between parallel and
antiparallel cases.  At large separations, $M_{tot}$ seems to tend to
$M_{tot}^*$ taking into account that $M_{tot}$ computed at $r=512$ is
slightly less than $M_{tot}$ at $r=\infty$.  Again, we conclude from
figures \ref{fig:14A} and \ref{fig:14B} that both conformal-imaging
and puncture solutions contain junk fields. The amount of junk appears
to be less in the puncture method.

\begin{figure}
\begin{center}
\epsfig{file=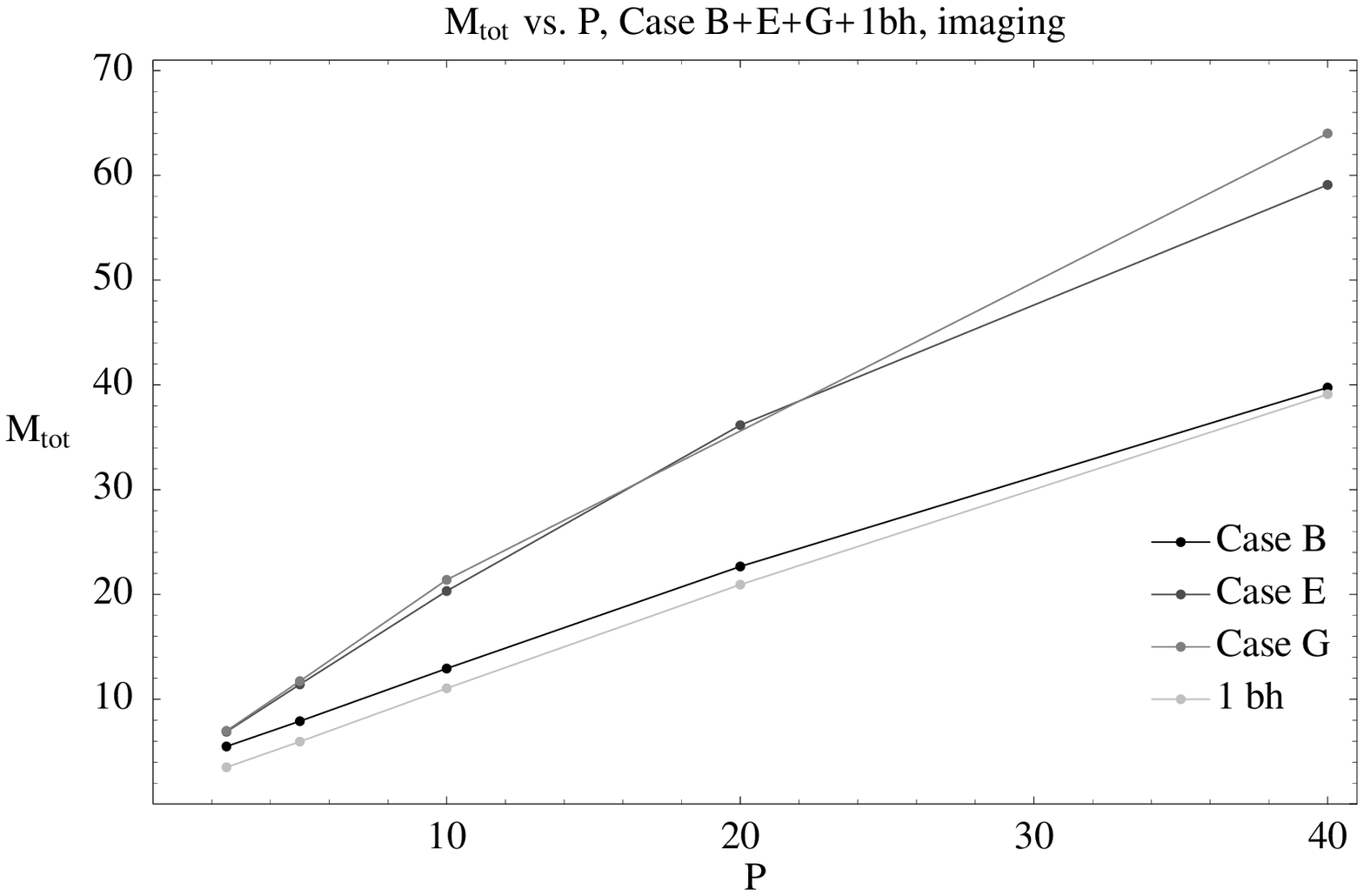,width=17cm}
\caption[]{Total mass, $M_{tot}$ as a function of linear momentum, $P$, 
for separation $d=128$ for two black holes with configurations B, E, G 
(figure \ref{fig:11}), imaging method 
\label{fig:15A}}
\end{center}
\end{figure}

\begin{figure}
\begin{center}
\epsfig{file=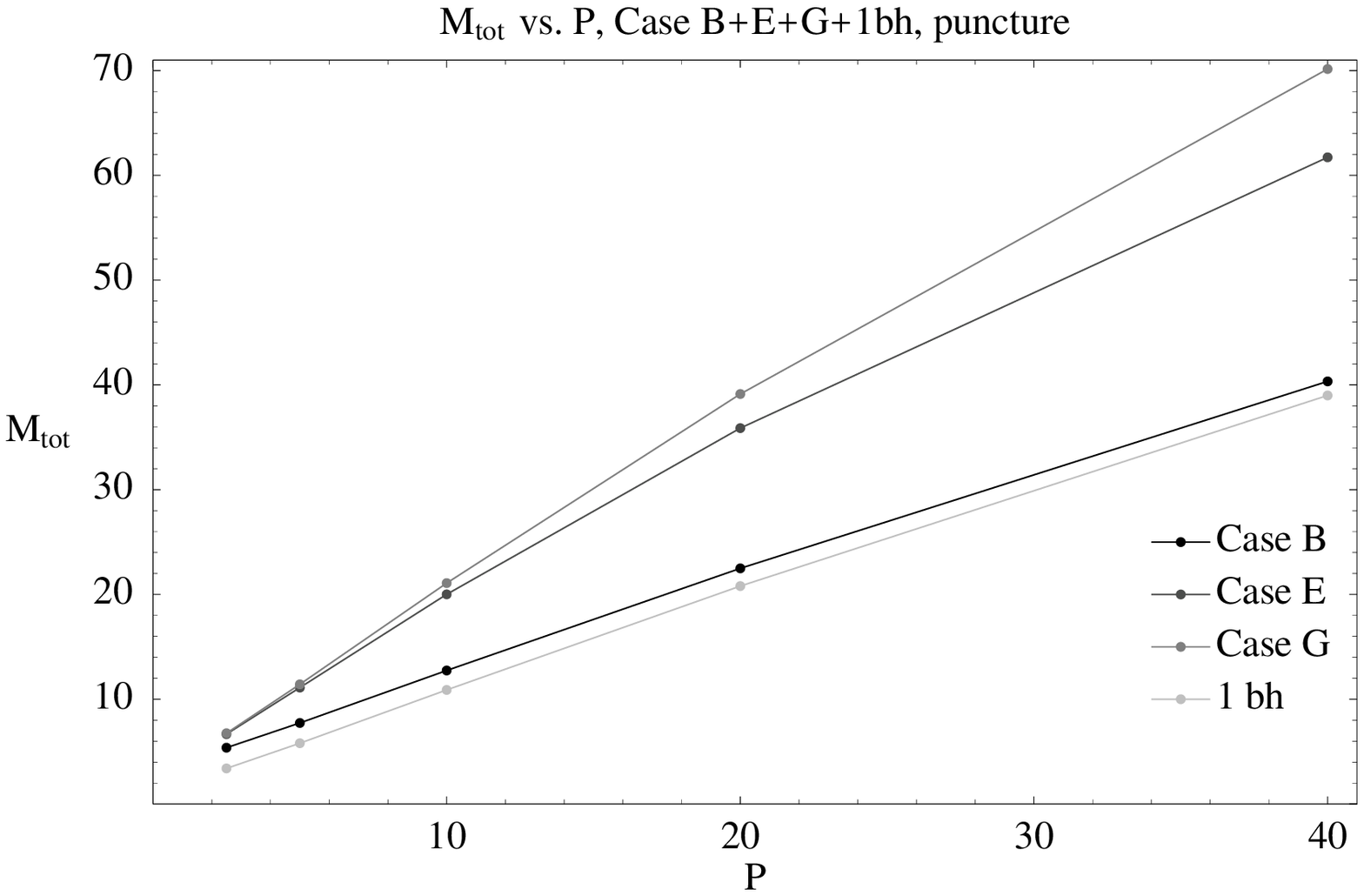,width=17cm}
\caption[]{Total mass, $M_{tot}$ as a function of linear momentum, $P$, 
for separation $d=128$ for two black holes with configurations B, E, G 
(figure \ref{fig:11}), puncture method
\label{fig:15B}}
\end{center}
\end{figure}

Figures \ref{fig:15A} and \ref{fig:15B} show $M_{tot}$ as a function
of $P$ for black hole configuration B with only one black hole having
non-zero $P=5$ and for configurations E and G with anti-parallel and
parallel linear momenta $P=5$ for the separation $d=128$ between the
black holes. We see that $M_{tot}$ is somewhat larger for black holes
with parallel momenta than for black holes with anti-parallel
momenta. This is the same tendency which we already saw on figures
\ref{fig:14A} and \ref{fig:14B}.  We found that the values of
$M_{tot}$ for cases E and G are somewhat smaller that the sum of
$M_{tot}$ of single black hole with the same values of $P$.  For case
B, $M_{tot}$ is very close to the sum of $M_{tot}$ a Schwarzschild
black hole with $P=0$ and a boosted black hole with linear momentum
$P=5$.

\subsection{\bf \normalfont \normalsize 
\it Two black holes with angular momenta}

\begin{figure}
\begin{center}
\epsfig{file=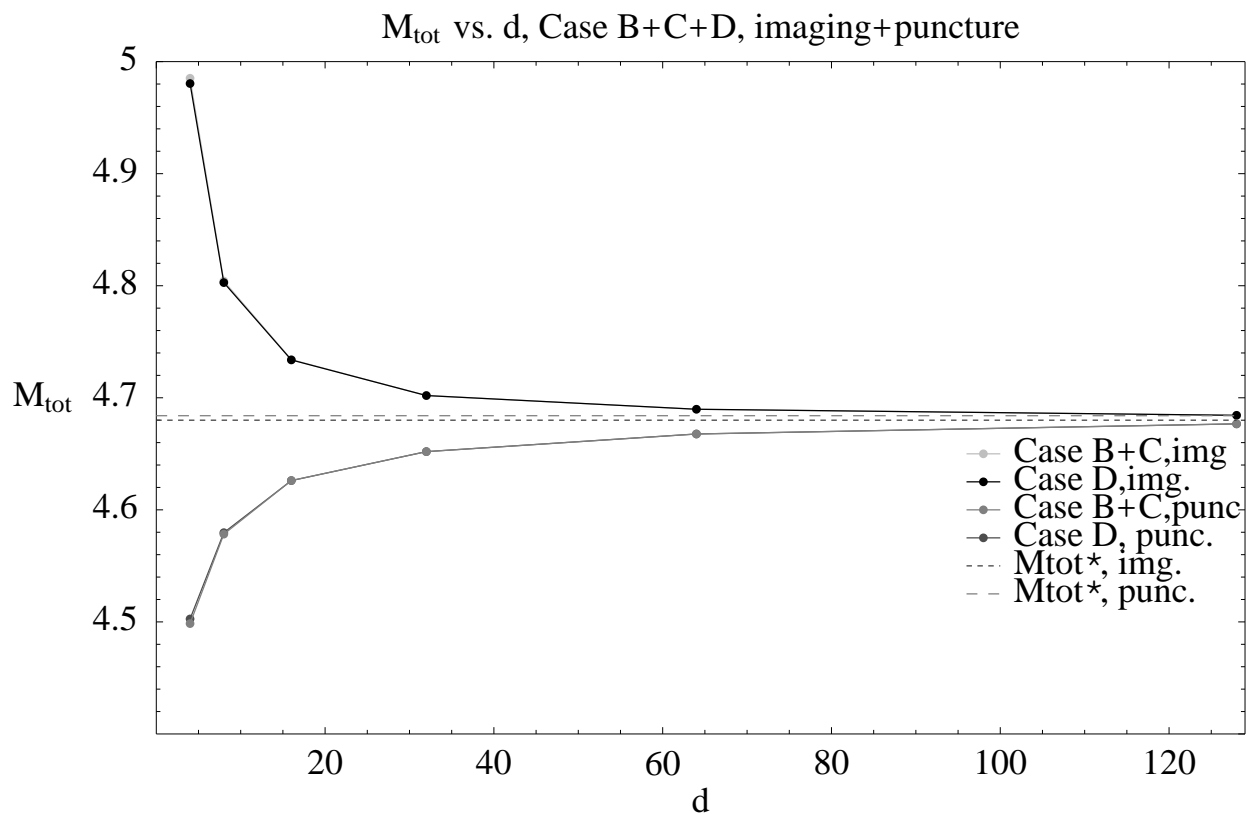,width=17cm}
\caption[]{Total mass, $M_{tot}$, as a function of separation, $d$, 
for two black holes with configurations B, C, D (figure \ref{fig:11}) with
angular momentum
\label{fig:16}}
\end{center}
\end{figure}

Figure \ref{fig:16} shows $M_{tot}$ as a function of separation $d$
for configurations with two black holes where one black hole has $S=0$
and the other one has $S=5$, calculated by both conformal-imaging and
puncture methods. First, Figure \ref{fig:16} shows that $M_{tot}$ does
not depend on the relative orientation of $\bf S$ with respect to the
line connecting the black hole throats or punctures.  At large
separations $M_{tot}$ tends to the sum of the numerical $M_{tot}$ of
two single black holes with $M=2$ and the same $S=0$ and $S=5$,
respectively. As it was for similar configurations with linear momenta
(see figures \ref{fig:13A} and \ref{fig:13B}), the amount of junk is
greater in conformal-imaging solutions.

\begin{figure}
\begin{center}
\epsfig{file=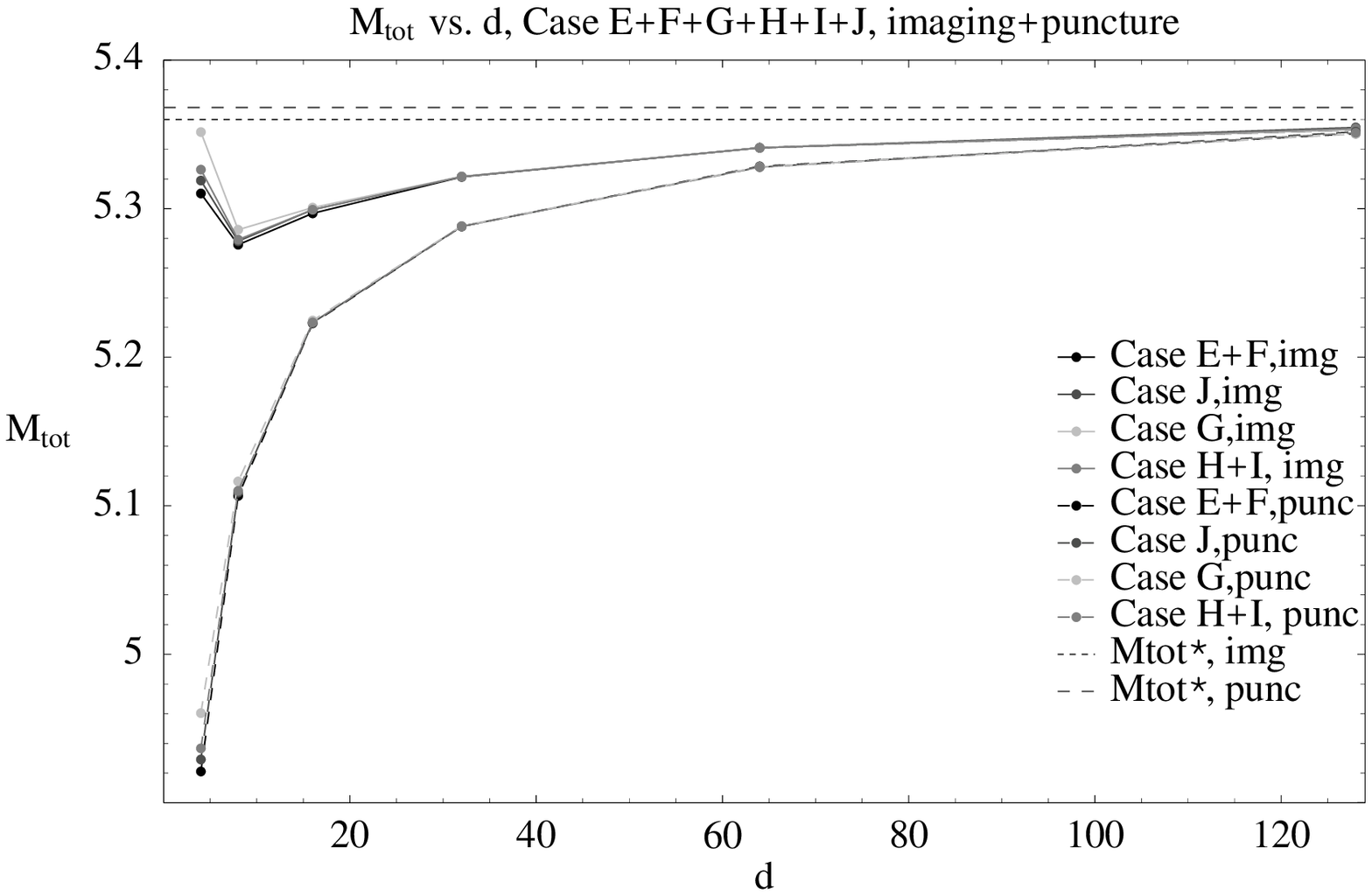,width=17cm}
\caption[]{Total mass, $M_{tot}$, as a function of separation, $d$, 
for two black holes with configurations  E, F, G, H, I, J 
(figure \ref{fig:11}) with angular momentum
\label{fig:17}}
\end{center}
\end{figure}

Figure \ref{fig:17} shows $M_{tot}$ as a function of separation $d$
for two black hole configurations with both black holes having
non-zero angular momentum $S=5$, calculated by both conformal-imaging
and puncture methods. Again, $M_{tot}$ does not depend on relative
orientation of angular momenta. At large $d$, $M_{tot}$ tends to the
sum of $M_{tot}$ for two individual rotating black hole with $S=5$.
Again, the amount of junk is greater in conformal-imaging solution.

\begin{figure}
\begin{center}
\epsfig{file=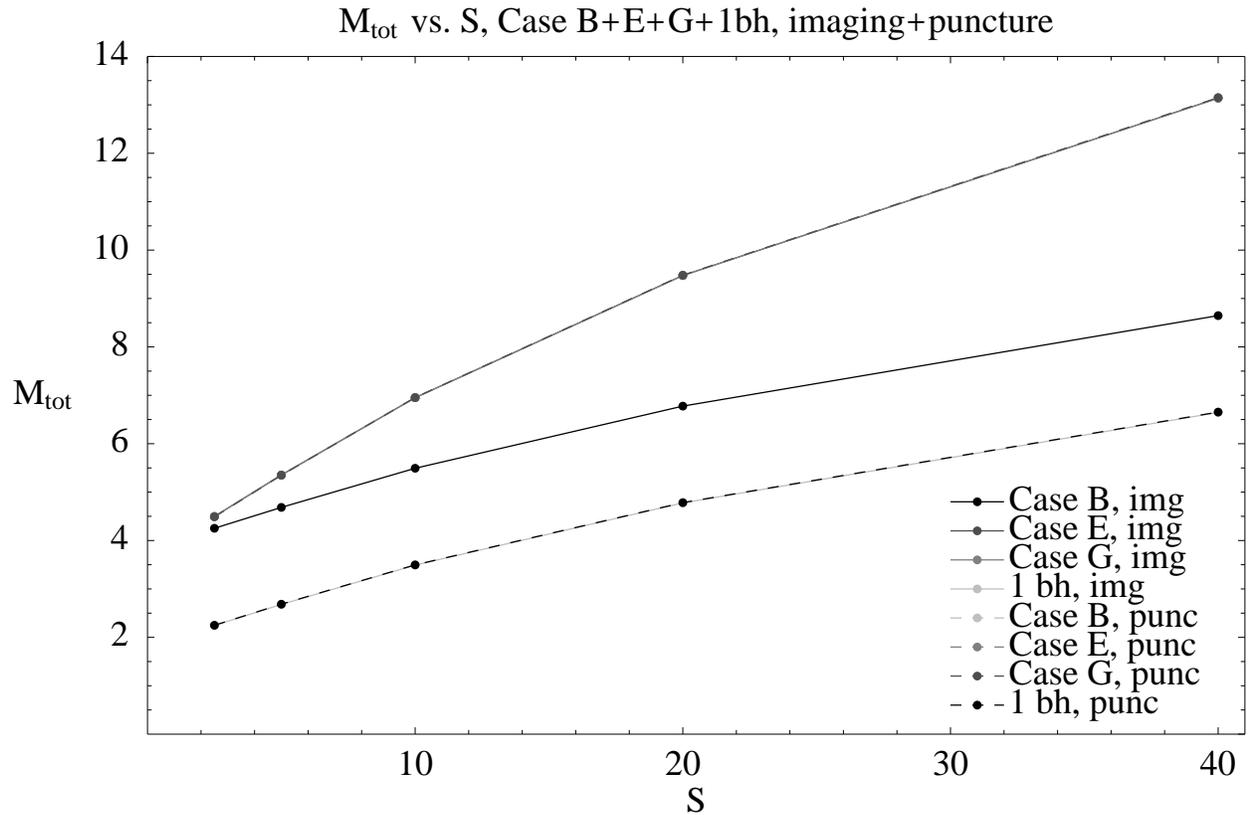,width=17cm}
\caption[]{Total mass, $M_{tot}$ as a function of angular momentum, $S$, 
for separation $d=128$ for two black holes with configurations B, E, G 
(figure \ref{fig:11})
\label{fig:18}}
\end{center}
\end{figure}

Figure \ref{fig:18} shows $M_{tot}$ as a function of $S$ for all
two-black hole configuration with angular momenta (Figure
\ref{fig:11}) and separation $d=128$, computed by both imaging and
puncture methods. We found that at this separation, $M_{tot}$ for all
of these configurations practically coincide with the sum of $M_{tot}$
of two individual black holes with the corresponding momenta.

\section{Conclusions}
\label{sec:6}
\setlength{\parindent}{0em} \setlength{\parskip}{1.5ex plus 0.5ex}

We calculated conformal initial value data for single and binary black
hole configurations on a high-resolution, adaptively refined Cartesian
mesh using both conformal-imaging and conformal-puncture
methods. Adaptive mesh refinement approach allowed us to obtain
accurate three-dimensional numerical solutions both near and far away
from black hole for a wide range of black hole configurations with
various linear and angular momenta and various black hole
separations. The equivalent uniform-grid resolution obtained in the
calculations was $65536^3$.

Determination of the total mass $M_{tot}$ of configurations required
an integration over a surface at infinity (\ref{eq:12}). Using numerical
values of $M_{tot}$ obtained by integration over a sphere at large
finite radii $r \leq 500 r_g$, where $r_g = 2M/c^2$ allowed us to
obtain $M_{tot}(\infty)$ by extrapolation.  For configurations with
large linear momenta, the difference between $M_{tot}(\infty)$ and
$M_{tot}(500r_g)$ was as large as $\simeq 0.1 M_{tot}$. The difference
was substantially smaller for configurations with large angular
momenta.  Comparison with known analytic solutions was then performed
using asymptotic values of $M_{tot}(\infty)$.

For a single rotating black hole, the mass of the calculated apparent
horizon $M_{ah}$ is less than the irreducible mass of a Kerr black
hole with the same $M_{tot}$. For a single boosted black hole, the
calculated $M_{ah}$ is less than that of the corresponding boosted
Schwarzschild black hole. This shows that the junk field is present in
the conformal solutions. Comparison of the spatial distribution of
local invariants for both rotating and boosted black hole shows that
junk field is present at large distances from the holes. For a
rotating black hole, the shape of the numerical apparent horizon is
much more spherical than that of a Kerr black hole. It appears the
conformal approach generates more spherically-symmetric solutions.

For two-black hole configurations, with increasing separation $d$
between the black holes the value of $M_{tot}(d)$ tends to the sum of
$M_{tot}$ of two individual single-black holes numerical solutions
containing their corresponding junk fields. If only one of the black
holes has a non-zero linear or angular momentum, the value of
$M_{tot}$ is practically independent of the orientation of the
momentum.  For black holed both having non-zero angular momentum,
$M_{tot}$ is also practically independent of the orientation of the
momenta. For black hole both having non-zero linear momenta, the value
of $M_{tot}$ is maximal for parallel momenta and minimal for
anti-parallel momenta.  In general, the conformal-puncture method
appears to generate less junk gravitational field than the
conformal-imaging method.

This work confirms that conformal initial conditions are not suitable
as initial conditions for astrophysical black hole collisions since
they contain junk gravitational field. They are useful as initial
conditions for testing time-integration schemes for black hole
collisions because black holes can be places at small initial
separation.  Our adaptive mesh refinement approach allows to construct
these initial data on a Cartesian mesh with a high accuracy. We plane
to use similar adaptive mesh refinement for the integration of these
initial data in time.

\section{Acknowledgments}
\label{sec:7}
\setlength{\parindent}{0em} \setlength{\parskip}{1.5ex plus 0.5ex}

We thank Kip Thorne and members of his relativity group for useful
discussions. N. Jansen thanks the numerical relativity group at the
Max Planck Institut f$\ddot{u}$r Gravitationsphysik, Albert Einstein
Institut for useful discussions. I. Novikov thanks the NRL for
hospitality during his stay.  The work was supported in part by the
NASA Space Astrophysics grant SPA-00-067, Office of Naval Research,
Danish Natural Science Research Council grant No.9401635, and by
Danmarks Grundforskningsfond through its support for the establishment
of the Theoretical Astrophysics Center.

\appendix
\section{Appendix: Proof}
\label{ap:A}
{\bf Proposition:} Let a configuration of two black holes be
given. Suppose that the linear momentum parameters of the problem are
$P1$ and $P2$, and that the angular momentum parameters for both holes
are the null-vector. Then the metric that is a solution to the
constraint equations with these parameters is the same as the metric
of that is the solution to the constraint equations where the linear
momentum parameters are $-P1$ and $-P2$, i.e. the problem where the
sign of all components of the linear momentum parameters has been
reversed.

{\bf Proof:}
The energy constraint is given by:
\begin{equation}
\nabla^2 \phi + \frac{1}{8} \phi^{-7} K_{ij} K^{ij} = 0 \quad, 
\end{equation}

As we can see, the extrinsic curvature enters only as $K_{ij}
K^{ij}$. Thus, to prove the proposition, we need to prove that this
term does not depend on the sign of the linear momentum vectors. In
the puncture formulation the extrinsic curvature tensor of the
configuration $P1,P2$ is given by:
\begin{eqnarray}
K^{ij} &=& \frac{3}{2 r1} \left(P1^i n1^j + P1^j n1^i - \left( f^{ij} -
n1^i n1^j\right) P1^l n1_l\right) +\nonumber \\ 
&\quad& \frac{3}{2 r2} \left(P2^i n2^j + P2^j n2^i - \left( f^{ij} -
n2^i n2^j\right) P2^l n2_l\right)
\end{eqnarray}
Trivial manipulations give that:
\begin{eqnarray*}
K_{ij} K^{ij} &=& \frac{9}{4}\,\left(\frac{1}{r1^4}\left( n1_i\,n1^j\,
P1_j\,P1^i + n1_i\,n1^i\,P1_j\,P1^j - 
2\,n1_c\,n1^i\,P1_i\,P1^c  \right.\right.\\
&&\left.\left.
+ n1_i\,n1_c\,
n1^i\,n1^j\,P1_j\,P1^c -
2\,n1_i\,
n1_c\,P1^i\,P1^c + 3\,(n1_c)^2\,
(P1^c)^2 -  \right.\right.\\
&&\left.\left.
2\,n1_i\,(n1_c)^2\,n1^i\,
(P1^c)^2 + n1_j\,n1^i\,P1^j\,
\left( P1_i + n1_i\,n1_c\,P1^c \right) + 
 \right.\right.\\
&&\left.\left.
n1_j\,n1^j\,\left( P1_i + n1_i\,n1_c\,
P1^c \right) \,\left(P1^i + n1_c\,n1^i\,
P1^c \right)\right) \right. \\
&&\left. + \frac{1}{r1^2\,r2^2} \left(n1^j\,
n2_j\,P1^i\,P2_i + n1^i\,n2_j\,
P1^j\,P2_i +
n1_c\,n1^i\,n1^j\,
n2_j\,P1^c\,P2_i 
 \right.\right.\\
&&\left.\left.
- 2\,n1_c\,n2^i\,
P1^c\,P2_i + n1^j\,n2_i\,P1^i\,
P2_j + n1^i\,n2_i\,P1^j\,P2_j +  \right.\right.\\
&&\left.\left.
n1_c\,n1^i\,n1^j\,n2_i\,P1^c\,P2_j +
n1_i\,n2^j\,P1_j\,P2^i - 2\,n1_c\,
n2_i\,P1^c\,P2^i +  \right.\right.\\
&&\left.\left. n1_j\,n2^j\,
\left( P1_i + n1_i\,n1_c\,P1^c \right) \,P2^i + \,
n1_i\,n2^i\,P1_j\,P2^j + \right.\right.\\
&&\left.\left. 
n1_j\,n2^i\,\left( P1_i + n1_i\,n1_c\,P1^c \right) 
P2^j 
- 2\,n1^i\,n2_c\,P1_i\,P2^c + n1_i\,
n2_c\,n2^i\,n2^j\,P1_j\,P2^c - \right.\right.\\
&&\left.\left. 
2\,n1_i\,n2_c\,
P1^i\,P2^c + n1^j\,n2_i\,n2_j\,n2_c\,P1^i\,
P2^c + n1^i\,n2_i\,n2_j\,n2_c\,P1^j\,
P2^c \right.\right.\\
&&\left.\left. 
+ 6\,n1_c\,n2_c\,P1^c\,P2^c - 
2\,n1_i\,n1_c\,n1^i\,n2_c\,P1^c\,P2^c + n1_c\,
n1^i\,n1^j\,n2_i\,n2_j\,n2_c\,P1^c\,
P2^c - 
\right.\right.\\
&&\left.\left. 
2\,n1_c\,n2_i\,n2_c\,n2^i\,
P1^c\,P2^c +  n1_j\,n2_c\,n2^i\,n2^j\,
\left( P1_i + n1_i\,n1_c\,P1^c \right) 
P2^c\right) \right.\\
&&\left.
+\frac{1}{r1^4} \left(n2_j\,
n2^j\,P2_i\,P2^i + n2_i\,n2^j\,P2_j\,
P2^i + 
n2_j\,n2^i\,P2_i\,P2^j + n2_i\,n2^i\,
P2_j\,P2^j 
\right.\right.\\
&&\left.\left.
- 2\,n2_c\,n2^i\,P2_i\,P2^c +
n2_j\,n2_c\,n2^i\,n2^j\,P2_i\,P2^c + 
n2_i\,n2_c\,
n2^i\,n2^j\,P2_j\,P2^c 
\right.\right.\\
&&\left.\left.
- 2\,n2_i\,n2_c\,P2^i\,
P2^c + 
n2_i\,n2_j\,n2_c\,n2^j\,P2^i\,P2^c + 
n2_i\,n2_j\,n2_c\,n2^i\,P2^j\,P2^c +  \right.\right.\\
&&\left.\left.
3\,(n2_c)^2\,(P2^c)^2 - 2\,n2_i\,(n2_c)^2\,
n2^i\,(P2^c)^2 + n2_i\,n2_j\,(n2_c)^2\,
n2^i\,n2^j\,(P2^c)^2\right)\right)
\end{eqnarray*}
What's interesting about this equation is that each term have the
product of two components of the momentum parameter vectors. Thus, if
we change the sign on all components of the momentum parameter
vectors, it will not change anything in the value of the
$K_{ij}K^{ij}$ term, and thus the metric is conserved under such a
sign change, at least in the puncture formulation. In the Cook
formulation, there are more terms in the expression for $K_{ij}$ and
$K^{ij}$ but each of these terms is a product of some factor that
depends on the position where we wish to find $K_{ij}$ and of the
extrinsic curvature tensor, evaluated at some different position. This
means, that we multiply two terms where the normal-vector and the the
radius that enters in the expression for the extrinsic curvature
tensor are different, but the parameters P1 and P2 does not
change. Thus each term in the product $K_{ij}K^{ij}$ will contain a
product of two terms of the momentum vector parameters, and thus the
whole expression is conserved under the sign change of $P_1,P_2$. QED.

\clearpage
\newpage

\end{document}